\documentclass[%
 reprint,
 amsmath,amssymb,
 aps,
prd,
]{revtex4-2}

\usepackage{graphicx}% Include figure files
\usepackage{dcolumn}% Align table columns on decimal point
\usepackage{bm}% bold math
\usepackage{tensor}
\usepackage{color}
\usepackage{ulem}
\usepackage{orcidlink}

\usepackage{natbib}
\usepackage{hyperref}% add hypertext capabilities
\newcommand\numberthis{\addtocounter{equation}{1}\tag{\theequation}}
\DeclareMathOperator{\e}{e}

\begin{document}

\preprint{APS/123-QED}

\title{Chains of Rotating Boson Stars}

\author{Romain Gervalle \orcidlink{0000-0002-8978-2760}}
 \email{romain.gervalle@lmpt.univ-tours.fr}
\affiliation{
    Institut Denis Poisson, UMR - CNRS 7013,\\
    Universit{\'e} de Tours, Parc de Grandmont, 37200 Tours, France
}
\def\ls{\textcolor{blue}}

\date{\today}

\begin{abstract}
    Boson Stars are stationary, axially symmetric solutions of a complex scalar field theory coupled to gravity. Recently, multi-solitonic configurations interpreted as static \textit{chains} of multiple Boson Stars bound by gravity and carrying no angular momentum were reported. We propose to generalize those solutions to the stationary case by constructing chains of rotating Boson Stars and analyze their properties. The non-linear elliptic field equations are solved using the finite element method. We find that chains with an \textit{even} number of constituents exhibit the same spiral-like frequency dependence of their mass, angular momentum and Noether charge as single Boson Stars. In contrast, sequences of chains with an \textit{odd} number of constituents show nontrivial loops starting and ending at the flat vacuum. As a consequence, such solutions cannot be uniquely parametrized by a single parameter. We conjecture that all rotating chains correspond to excitations of single Boson Stars or pairs of them. We also analyze their flat space limit and find that they reduce to chains of $Q$-balls.
\end{abstract}

\maketitle

\section{Introduction}

Non-topological solitons were first established by \citet{Rosen1968,Friedberg1976} and since have been a topic of great interest in field theory. They consist in non-trivial stationary and spatially localized solutions of a non-linear physical theory in flat space which have a finite energy. In contrast to topological solitons, their existence is guaranteed by the conservation of a Noether charge rather than a topological charge. A simple example of these solitons are \textit{$Q$-balls}, introduced by \citet{Coleman1985}, which are non-topological solitons in a complex scalar field theory with a harmonic time dependence and a suitably chosen self-interacting potential. This theory presents a global phase invariance which provides the conserved charge $Q$. The complex field corresponds to interacting massive bosons. In light of this, $Q$-balls can be seen as an agglomerate of massive bosons in flat space tied together by their self-interactions and the charge $Q$ is related to the number of particles \cite{Friedberg1976,Coleman1985}. The minimization of the energy of a multi-particle system with conserved charge $Q$ favors a configuration in which particles are glued together in a soliton, rather than a configuration with individual free particles. This ensures the existence of $Q$-balls \cite{Coleman1985}.

For spherically symmetric $Q$-balls, the complex field may be written as $\Phi=\phi(r)\e^{i\omega t}$ with $\phi$ a real function of the radial coordinate, $\omega$ a constant frequency and $t$ the time coordinate. The harmonic time dependence is a crucial ingredient which, together with a well chosen potential, allows the existence of solitons \cite{Derrick1964}. Because of the spherical symmetry, surfaces of constant energy density are spheres. Solutions exist in a finite frequency range $\omega_\text{min}<\omega<\omega_\text{max}$ : the upper bound corresponds to the mass of the scalar field whereas the lower bound depends on the shape of the potential \cite{Friedberg1976,Coleman1985,Lee1992a}. The mass $M$ of a $Q$-ball and its charge $Q$ reach their minimal values at a critical frequency $\omega_\text{cr}\in(\omega_\text{min},\omega_\text{max})$ and they both diverge at the boundaries of the frequency domain. For a fixed frequency $\omega$, there are infinitely many spherically symmetric $Q$-balls labeled by the number $n$ of nodes of their $\phi$-profile \cite{Volkov2002a}. Solutions with $n\geq 1$ are called radially excited $Q$-balls and are unstable, whereas solutions with $n=0$ are referred to as fundamental $Q$-balls and are stable in the range $\omega_\text{min}<\omega<\omega_\text{cr}$ \cite{Friedberg1976}. In general, finding $Q$-ball solutions requires to solve a non-linear differential equation numerically, however, it has been found recently that very accurate analytical approximations of stable $Q$-balls can be obtained \cite{Heeck2021}.

Rotating generalizations of $Q$-balls exist \cite{Volkov2002a,Kleihaus2005}. The rotation is achieved by including an additional phase factor $\e^{im\varphi}$ to the scalar field, where $\varphi$ is the azimuthal angle and $m$ is the integer rotational number. Rotating solutions are classified according to their behavior under reflections with respect to the equatorial plane. The energy density isosurfaces of such spinning $Q$-balls consist in one or more tori and the field configuration carries an angular momentum $J$ directly related to the charge by $J=m\,Q$. Their angular momentum is thus quantized at the classical level and as a consequence, for a fixed $Q$, there are no slowly rotating $Q$-balls (with arbitrarily small $J$). Apart from a different spatial profile and a non-zero angular momentum, spinning $Q$-balls share the same properties as the spherically symmetric ones regarding the frequency dependence of their mass or charge.

\textit{Boson Stars} (BSs) arise when $Q$-balls are coupled to gravity \cite{Lee1992a,Friedberg1987,Mielke1998}. They represent solitonic configurations with globally regular geometry (no horizon nor singularities). Because of the attractive nature of gravitation, BSs exist also if the scalar field potential contains no self-interaction terms hence only a mass term \cite{Kaup1968,Ruffini1969}. In this case, the solutions are often called \textit{mini}-Boson Stars because their maximum mass is very small, except for tiny values of the boson mass \cite{Jetzer1992}. BSs can carry angular momentum (like spinning $Q$-balls) \cite{Schunck1996,Kleihaus2005,Kleihaus2008} and obey the quantization relation $J=m\,Q$. Although the spacetime geometry of BSs is regular everywhere, ergoregions may appear for rotating BSs \cite{Kleihaus2008,Cardoso2008}. It is also worth mentioning that self-gravitating solitons exist also if the scalar field is replaced by a vector or a spinor field \cite{Herdeiro2017,Herdeiro2019,Herdeiro2020}.

Previous studies of BSs (for reviews, see Refs.~\cite{Jetzer1992,Lee1992a,Schunck2003,Liebling2017,Shnir2022}) were motivated by different elements. First, the discovery of the Higgs boson \cite{atlas2012,cms2012} has proven that fundamental scalar fields exist in Nature. In this sense, BSs can be seen as a toy-model for solitonic configurations of more realistic fields. Second, they can have application in astrophysics by mimicking black holes and thus avoiding the presence of an event horizon \cite{Guzman2009,Herdeiro2021c}. Third, scalar fields are also a fundamental ingredient in many cosmological models, for example primordial inflation models (\textit{inflaton}) \cite{Peter2013} or quintessence models of dark energy \cite{Tsujikawa2013}.

In the present work, we consider BSs which have flat space counterparts ($Q$-balls): the scalar field potential contains self-interactions carried by terms quartic and sextic in $|\Phi|$. Although the domain of existence of solutions is also a finite frequency range as for $Q$-balls, the presence of gravity has a crucial influence on the frequency dependence of the mass and the charge. Close to $\omega_\text{max}$, BSs emerge from the trivial vacuum at which $M$ and $Q$ vanish (rather than diverge in the $Q$-ball case). Constructing a sequence of fundamental BSs, the $(\omega,M)$ and $(\omega,Q)$ curves present a spiraling behavior so that the solutions are no longer uniquely determined by their scalar field frequency. The configurations tend to a limiting solution with finite mass and charge at the center of the spiral \cite{Friedberg1987}. For excited BSs \cite{Collodel2017}, the spiraling pattern occurs for non-rotating configurations. For even parity rotating solutions, the spiraling behavior is replaced by a loop pattern -- BSs sequences start and end at $\omega_\text{max}$ where they approach the vacuum configuration. For excited rotating BSs with odd parity, the frequency dependence pattern has not been studied yet.

The aim of this paper is to construct chains of rotating BSs and analyze their properties. In the non-rotating case, chains with two constituents were first considered by \citet{Yoshida1997} and then generalized to larger numbers of constituents in Ref.~\cite{Herdeiro2021}: they consist in static, axisymmetric equilibrium configurations interpreted as several BSs located along the symmetry axis which are tied together by the gravitational attraction and are kept apart from each other by a scalar repulsion \cite{Battye2000,Bowcock2009}. Including a non zero angular momentum to these solutions is a very natural generalization and we numerically construct the rotating chains. While the numerical scheme commonly employed in the literature is based on the finite difference method \cite{Schonauer1989,Schauder1992}, we use a different algorithm to solve the non-linear elliptic Partial Differential Equations (PDEs): the finite element method.

%(for mini-BS such configurations exist as well, see Ref.~\cite{Herdeiro2021a})
% First, we reproduce sequences of already known solutions -- rotating BSs with even/odd parity \cite{Kleihaus2005,Kleihaus2008} and non-rotating chains of BSs \cite{Herdeiro2021} -- to test our numerical solver. \ls{In Sect. I, we test our numerical solver on sequences of already know solutions: rotating BSs with even/odd parity \cite{Kleihaus2005,Kleihaus2008} and non-rotating chains of BSs \cite{Herdeiro2021}.} Then we construct \ls{the} rotating chains, starting from a suitable initial guess for the Newton-Raphson \ls{where is it said that youuse this method ?} iterations. \ls{Finally, we analyze their properties and find that chains with an odd number of constituents exhibit a loop pattern for their mass or charge frequency dependence. Such chains turn out to correspond exactly to radial excitations of single BSs previously obtained in \cite{Collodel2017} and thus are likely to be unstable. For chains with an even number of constituents, we find that the $\omega$-dependence of sequences shows a spiral behavior just as the fundamental rotating BSs. To our knowledge, the chains with a even number of constituents larger than two were never reported in the literature so far. We conjecture that they correspond to excitations of BS pairs. / this is conclusion material ! or at least should be in the section not here}

The rest of the paper is organized as follows. In Section~\ref{model}, we describe the model: recalling the action, the general field equations and the conserved quantities. We also give the ansatz for the axisymmetric fields and specify the boundary conditions. In Section~\ref{numerics}, we present our numerical approach based on the finite element solver FreeFem++ \cite{MR3043640}. In Section~\ref{solutions}, we reproduce sequences of already known solutions -- rotating BSs with even/odd parity \cite{Kleihaus2005,Kleihaus2008} and non-rotating chains of BSs \cite{Herdeiro2021} -- to test our numerical solver; we also construct new rotating solutions in this section and analyze their properties and their flat space limit. The Section~\ref{conclusion} gives our conclusions and perspectives. In the Appendix, we give the coupled set of elliptic PDEs to be solved. The metric signature is chosen to be $(-,+,+,+)$.

\section{The model}

\label{model}

\subsection{Action and conserved quantities}

We consider the theory of a complex scalar field $\Phi$ minimally coupled to Einstein's gravity. The (dimensionless \footnote{In the dimensionfull action, the potential $U(|\Phi|^2$) contains an overall factor that has been set to unity by a rescaling.}) action is given by
\begin{widetext}
\begin{equation}
    \label{action}
    S=\int{d^4 x\sqrt{-g}\left(\frac{R}{4\alpha^2}-\frac{1}{2}g^{\mu\nu}\left(\partial_\mu\Phi^\ast\partial_\nu\Phi+\partial_\nu\Phi^\ast\partial_\mu\Phi\right)-U(|\Phi|^2)\right)},
\end{equation}
\end{widetext}
where $R$ is the Ricci scalar of the spacetime metric $g_{\mu\nu}$, $\alpha$ is the gravitational coupling and $U$ is the scalar field potential. The simplest BSs can be obtained with a potential containing only a mass term \cite{Kaup1968,Ruffini1969} but here we consider a self-interacting potential
\begin{equation}
    \label{pot}
    U(|\Phi|^2)=|\Phi|^6-\lambda|\Phi|^4+u_0^2|\Phi|^2,
\end{equation}
where $u_0$ is the mass of the scalar excitations around the vacuum (boson mass) and $\lambda>0$ is a parameter determining the self-interactions.

If we remove gravitational interactions and fix the background metric to be Minkowski, the theory describes the so-called $Q$-balls, intensively studied in the literature \cite{Friedberg1976,Coleman1985,Lee1992a,Volkov2002a,Kleihaus2005,Kleihaus2008,Brihaye2008,Radu2008,Brihaye2009}. In this sense, BSs contained in the theory \eqref{action} with the sextic potential \eqref{pot} can be seen as gravitating $Q$-balls. The values of parameters in the potential are those commonly employed in the literature (see for example Refs.~\cite{Volkov2002a,Kleihaus2005,Kleihaus2008})
\begin{equation}
    u_0^2=1.1,\quad\quad\lambda=2.
\end{equation}
With these values, the potential has a global minimum at $|\Phi|=0$ for which it vanishes and a local minimum at some finite value of $|\Phi|$.

Varying the action \eqref{action} gives the field equations which consist in a non-linear Klein-Gordon equation
\begin{equation}
    \label{eqPhi}
    \left(\nabla_\mu\nabla^\mu-\frac{dU}{d|\Phi|^2}\right)\Phi=0,
\end{equation}
together with the Einstein equation
\begin{equation}
    \label{eineq}
    E_{\mu\nu}\equiv R_{\mu\nu}-\frac{1}{2}g_{\mu\nu}R-2\alpha^2 T_{\mu\nu}=0,
\end{equation}
where $T_{\mu\nu}$ is the stress-energy tensor which reads
\begin{align*}
    T_{\mu\nu}=&\partial_\mu\Phi^\ast\partial_\nu\Phi+\partial_\nu\Phi^\ast\partial_\mu\Phi\\&-g_{\mu\nu}\left(\frac{1}{2}g^{\alpha\beta}\left(\partial_\alpha\Phi^\ast\partial_\beta\Phi+\partial_\beta\Phi^\ast\partial_\alpha\Phi\right)+U(|\Phi|^2)\right).\numberthis{}
\end{align*}

We are interested in stationary axisymmetric solutions, therefore the spacetime possesses two Killing vector fields which are, in adapted coordinates
\begin{equation}
\label{killing}
    \xi=\partial_t,\quad\quad\chi=\partial_\varphi,
\end{equation}
where $t$, $\varphi$ are respectively the asymptotic time and azimuthal coordinates. We also assume that spacetime is asymptotically flat and therefore, the mass $M$ and angular momentum $J$ of the solutions can be obtained from the Komar expressions \cite{Wald1984}
\begin{align*}
    M&=\frac{1}{\alpha^2}\int_\Sigma{R_{\mu\nu}n^\mu\xi^\nu dV}\\
    \label{globalquantM}
    &=2\int_\Sigma{\left(T_{\mu\nu}-\frac{1}{2}g_{\mu\nu}T\right)n^\mu\xi^\nu dV},\numberthis{} \\
    J&=-\frac{1}{2\alpha^2}\int_\Sigma{R_{\mu\nu}n^\mu\chi^\nu dV}\\
    \label{globalquantJ}
    &=-\int_\Sigma{\left(T_{\mu\nu}-\frac{1}{2}g_{\mu\nu}T\right)n^\mu\chi^\nu dV},\numberthis{}
\end{align*}
where $\Sigma$ is a spacelike hypersurface, $n^\mu$ is the normal vector to $\Sigma$ such that $n_\mu n^\mu=-1$ and we have used the Einstein equation \eqref{eineq} in the second equalities to replace the Ricci tensor $R_{\mu\nu}$ by the stress-energy tensor.

In addition, the theory \eqref{action} possesses a global $U(1)$ symmetry corresponding to the invariance under transformations $\Phi\rightarrow\Phi\e^{i\beta}$, with a constant $\beta$. The 4-current associated to this symmetry is
\begin{equation}
    j_\mu=-i\left(\Phi\partial_\mu\Phi^\ast-\Phi^\ast\partial_\mu\Phi\right),\quad \nabla_\mu j^\mu=0,
\end{equation}
and integration over a spacelike hypersurface $\Sigma$ of its timelike component gives the conserved Noether charge
\begin{equation}
\label{chargecov}
    Q=\int_\Sigma{j^\mu n_\mu dV}.
\end{equation}

\subsection{Ansatz and boundary conditions}

In a system of adapted coordinates for which the Killing vectors are given by Eq. \eqref{killing}, the stationary axisymmetric spacetime metric is independent of $t$, $\varphi$ and the line element can be put in the Lewis-Papapetrou form~\cite{Kleihaus1998,Stephani2003}
\begin{align*}
    ds^2=&-f\,dt^2+\frac{\ell}{f}\Bigg(h\left(dr^2+r^2 d\theta^2\right)\\
    \label{ansatzmet}
    &+r^2\sin^2\theta\left(d\varphi-\frac{w}{r}dt\right)^2\Bigg),\numberthis{}
\end{align*}
where the four functions $f$, $\ell$, $h$ and $w$ depend on the quasi-isotropic spherical coordinates $(r,\theta)$. The symmetry axis of spacetime is the set of points such that the norm $\chi^\mu\chi_\mu=r^2\sin^2\theta\,\ell/f$ is vanishing. It corresponds to $\theta=0$ or $\pi$ (the $z$-axis in cylindrical coordinates). The Minkowski spacetime is recovered when $f=\ell=h=1$ and $w=0$.

For the scalar field $\Phi$ we adopt the stationary ansatz commonly used in the $Q$-ball/BS literature \cite{Schunck1996}
\begin{equation}
    \label{ansatzphi}
    \Phi=\phi(r,\theta)\e^{i\omega t+im \varphi},
\end{equation}
where $\phi$ is a real function, $\omega$ is the constant frequency parameter and $m$ is the constant rotational number. The latter has to be an integer to ensure the single-valuedness of the scalar field. We will assume without loss of generality that $m$ is positive. Note that the harmonic time and azimuthal dependences disappear at the level of the stress-energy tensor and factorize in the field equations. As pointed out in the introduction, solitonic solutions to the field equations will exist in a finite frequency range. In the $Q$-ball case, the boundaries of this range are completely determined by the shape of the potential \cite{Volkov2002a} whereas for BSs, the lower bound will also depends on the gravitational coupling $\alpha$ and the rotational number $m$.

With the line element \eqref{ansatzmet}, the normal vector in Eqs.~\eqref{globalquantM},\eqref{globalquantJ},\eqref{chargecov} is $n=\sqrt{f}dt$ and the volume element is $dV=\left(1/\sqrt{f}\right)\sqrt{-g}\,dr\,d\theta\,d\varphi$ so that the charge and mass are
\begin{widetext}
\begin{align}
    Q&=4\pi\int_0^\infty{dr\int_0^\pi{d\theta\,r^2\sin\theta\,\frac{\ell^{3/2}h}{f^2}\left(\omega+\frac{m\,w}{r}\right)\phi^2}},\\
    \label{globalM}
    M&=4\pi\int_0^\infty{dr\int_0^\pi{d\theta\,r^2\sin\theta\,\frac{\ell^{3/2}h}{f^2}\left(f\,U(\phi^2)-2\omega\left(\omega+\frac{m\,w}{r}\right)\phi^2\right)}},
\end{align}
\end{widetext}
while for the angular momentum, one finds the quantization relation
\begin{equation}    
    \label{globalJ}
    J=m\,Q,
\end{equation}
which was first derived in Ref.~\cite{Schunck1996}.

Alternatively, $M$ and $J$ can be read off the asymptotic expansions of the functions $f$ and $w$ \cite{Kleihaus2001}
\begin{equation} 
    \label{asympMJ}
    M=\frac{2\pi}{\alpha^2}\lim\limits_{r\rightarrow\infty}r^2\partial_r f,\quad\quad J=\frac{2\pi}{\alpha^2}\lim\limits_{r\rightarrow\infty}r^2 w.
\end{equation}
Since the computation of the mass and angular momentum from \eqref{globalM}-\eqref{globalJ} should agree with \eqref{asympMJ}, this provides a way to check the numerical accuracy of our solutions.

Now injecting \eqref{ansatzmet}-\eqref{ansatzphi} into the Einstein-Klein-Gordon equations \eqref{eqPhi}-\eqref{eineq} gives a coupled set of five PDEs for the unknown functions $\phi$, $f$, $\ell$, $h$, $w$ whose explicit expressions are given in the Appendix. The set of PDEs is elliptic and is solved as a boundary value problem with appropriate boundary conditions.

Because the theory depends only on the norm of the scalar field $|\Phi|$, the solutions are classified according to the behavior of the function $\phi$ under reflections with respect to the equatorial plane $\theta=\pi/2$,
\begin{align*}
    \mathcal{P}=1\quad\text{(even parity) :}&\quad\phi(r,\pi-\theta)=\phi(r,\theta),\\
    \mathcal{P}=-1\quad\text{(odd parity) :}&\quad\phi(r,\pi-\theta)=-\phi(r,\theta).\numberthis{}
\end{align*}
The geometry, however, is left invariant under these reflections,
\begin{align*}
    &f(r,\pi-\theta)=f(r,\theta),\;\ell(r,\pi-\theta)=\ell(r,\theta)\\
    &h(r,\pi-\theta)=h(r,\theta),\;w(r,\pi-\theta)=w(r,\theta).\numberthis{}
\end{align*}
This allows us to reduce the integration domain to $(r,\theta)\in[0,\infty)\times[0,\pi/2]$.

For $r\rightarrow\infty$ the metric is assumed to approach the Minkowski metric and the scalar field goes to its vacuum value. Therefore the boundary conditions at infinity are 
\begin{gather*}
    f\rvert_{r\rightarrow\infty}=1,\quad \ell\rvert_{r\rightarrow\infty}=1,\quad h\rvert_{r\rightarrow\infty}=1,\\
    w\rvert_{r\rightarrow\infty}=0,\quad\phi\rvert_{r\rightarrow\infty}=0.\numberthis{}
\end{gather*}
Then regularity of the solutions at origin requires that
\begin{equation}
    \partial_r f\rvert_{r=0}=0,\quad\partial_r \ell\rvert_{r=0}=0,\quad h\rvert_{r=0}=1,\quad w\rvert_{r=0}=0,
\end{equation}
while for the scalar field
\begin{align*}
    \partial_r\phi\rvert_{r=0}&=0\quad\text{if}\quad m=0\;\text{and}\;\mathcal{P}=1,\\
    \phi\rvert_{r=0}&=0\quad\text{otherwise}.\numberthis{}
\end{align*}
Reflection symmetry with respect to the equatorial plane requires that
\begin{gather*}
    \partial_\theta f\rvert_{\theta=\pi/2}=0,\quad\partial_\theta \ell\rvert_{\theta=\pi/2}=0,\\ \partial_\theta h\rvert_{\theta=\pi/2}=0,\quad\partial_\theta w\rvert_{\theta=\pi/2}=0,\numberthis{}
\end{gather*}
while the conditions for the scalar field depend on the parity,
\begin{align*}
    \partial_\theta\phi\rvert_{\theta=\pi/2}&=0\quad\text{if}\quad\mathcal{P}=1,\\
    \phi\rvert_{\theta=\pi/2}&=0\quad\text{if}\quad\mathcal{P}=-1.\numberthis{}
\end{align*}
Finally axial symmetry and regularity imply on the symmetry axis the conditions
\begin{equation}
    \partial_\theta f\rvert_{\theta=0}=0,\quad\partial_\theta \ell\rvert_{\theta=0}=0,\quad h\rvert_{\theta=0}=1,\quad\partial_\theta w\rvert_{\theta=0}=0,
\end{equation}
and for the scalar field
\begin{align*}
    \partial_\theta\phi\rvert_{\theta=0}&=0\quad\text{if}\quad m=0,\\
    \phi\rvert_{\theta=0}&=0\quad\text{if}\quad m\geq 1.\numberthis{}
\end{align*}
Furthermore, the absence of conical singularity also requires that $\partial_\theta h\rvert_{\theta=0}=0$. This additional constraint is not imposed in practice but we checked that it effectively holds after the resolution, up to numerical accuracy.

Let us also mention that the finite energy condition for the scalar field at infinity also gives us the maximal bound for the field frequency
\begin{equation}
\label{ommax}
    \omega\leq\omega_\text{max}=u_0,
\end{equation}
this ensures an asymptotically exponential falloff of the scalar field function.

\section{Numerical approach}

\label{numerics}

To construct rotating BSs and chains of them, the set of five coupled non-linear elliptic PDEs for the functions $(\phi,f,\ell,h,w)$ is solved numerically with the boundary conditions defined in the previous section. Numerical computations are performed using the finite element solver FreeFem++ \cite{MR3043640} together with the Newton-Raphson method to treat non-linear equations. There are very few examples in the literature of using the finite element method in General Relativity, \textit{e.g.} \citet{Zeng2021} use it in the context of BSs, but the software they employ is not mentioned. To our knowledge, it is the first time that the FreeFem++ solver is used for relativistic gravitational problems.

\begin{figure*}
    \centering
    \includegraphics[scale=0.65]{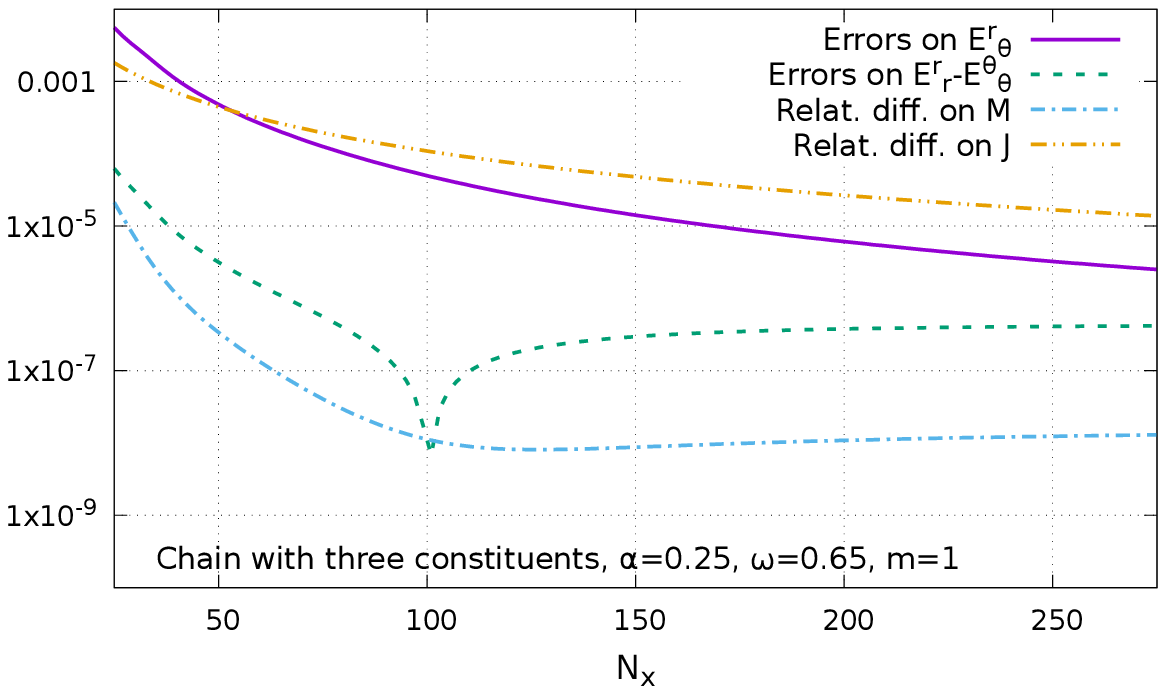}
    \includegraphics[scale=0.65]{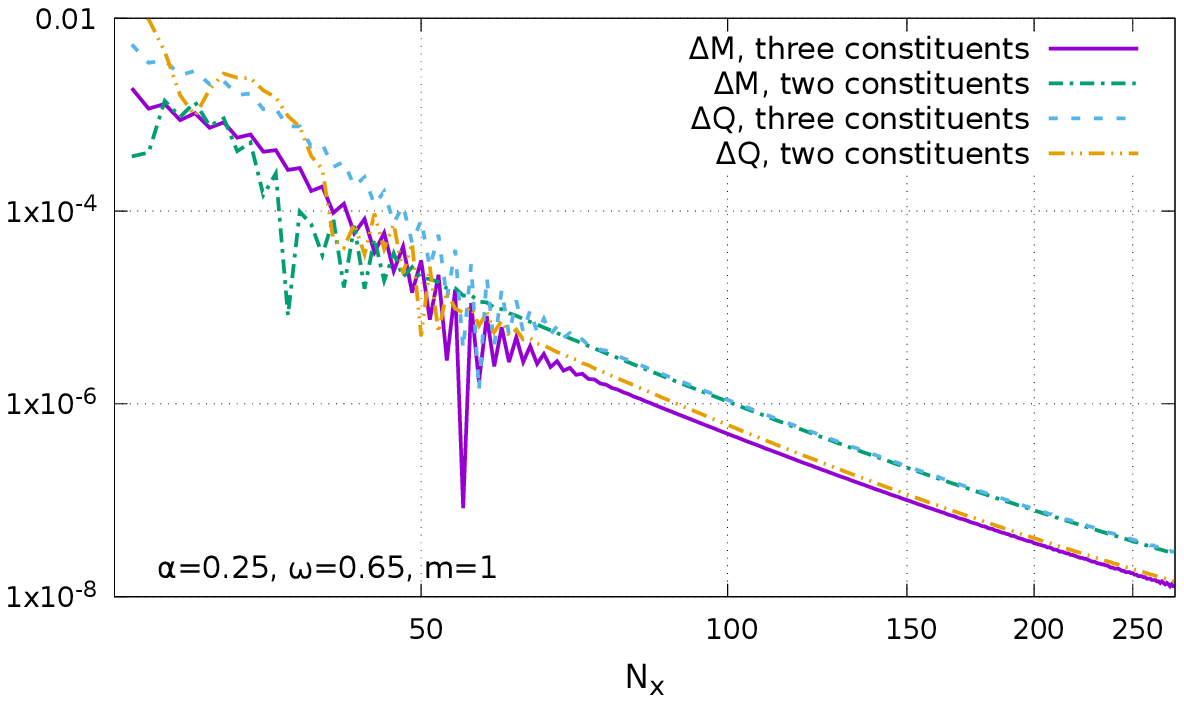}
    \caption{Left panel: Various error indicators against the number of points $N_x$ in the $x$-direction for a typical chain of BSs with three constituents. Right panel: Differences $\Delta M\equiv M(N_x+1)-M(N_x)$ and $\Delta Q\equiv Q(N_x+1)-Q(N_x)$ for typical chains with two and three constituents against $N_x$. Note that the peaks of the curves occur because of the logarithmic scale when the plotted quantities are coincidentally close to zero.}
    \label{err}
\end{figure*}

The Finite Element scheme requires equations to be in a weak form. 
As a first step, Eqs. \eqref{eqphi}-\eqref{eqh} are multiplied by $f/(r^2 lh)$ such that the second derivative terms are 
\begin{equation}
    \frac{f}{\ell h}\partial^2_{r}X +\frac{f}{r^2\ell h}\partial^2_{\theta} X=\left[\dots\right],
\end{equation}
with $X$ denoting the functions $\phi,f,\ell,h$ and $w$. This is the structure of the Laplace-Beltrami operator $\Delta\phi\equiv\frac{1}{\sqrt{-g}}\partial_\mu\left(\sqrt{-g}g^{\mu\nu}\partial_\nu\phi\right)$ for the metric given in eq.~\eqref{ansatzmet}. To avoid the use of a cutoff radius, we introduce a new compactified coordinate
\begin{equation}
    x\equiv\frac{r}{1+r},
\end{equation}
which maps the semi-infinite interval $r\in[0,\infty)$ to a finite one $x\in[0,1]$. Finally, the PDEs are multiplied by test functions, integrated over the whole domain, and the second derivatives are integrated by part.

The main advantage of the finite element scheme is that one is not restricted to a regular, rectangular grid for discretization. The FreeFem++ software has its own internal mesher which allows for irregular (triangular) meshes \cite{MR3043640}. Typically we choose a non-uniform repartition of triangles with a higher density close to infinity at $x=1$; however, depending on the solutions profiles, we can also choose a higher density close to the origin at $x=0$ or close to the $z$-axis at $\theta=0$.

The Einstein equation \eqref{eineq} contains two components, $E^r_\theta=0$, $E^r_r-E^\theta_\theta=0$, which are not solved in practice. Integrating them over a spacelike hypersurface gives an estimation of numerical errors. In addition, we also evaluate the relative difference on the mass and angular momentum computed from \eqref{globalM}-\eqref{globalJ} and from \eqref{asympMJ}. We have noticed that increasing the number of points $N_\theta$ in the $\theta$-direction does not change significantly the errors. Therefore, we fix $N_\theta=25$ and present the dependence of the error indicators on the number of points $N_x$ in the $x$-direction in left panel of Fig.~\ref{err}. In the rest of the paper, we choose $N_x=200$ so that our typical errors are at most of the order of $10^{-5}$ and the computation time for obtaining one solution on a personal computer with a parallelized code is about thirty seconds.

The right panel of Fig.~\ref{err} shows a convergence test of our code. We compute the mass and the charge for increasing values of $N_x$ and evaluate the differences $\Delta M\equiv M(N_x+1)-M(N_x)$, $\Delta Q\equiv Q(N_x+1)-Q(N_x)$. The latter are shown against $N_x$ on the figure with a logarithmic scale for both axes. After a oscillating phase when the number of points is too small, all the curves become straight lines with a slope of $-4$. This shows a fourth order convergence in the number of mesh points.

The input parameters of our code are the gravitational coupling $\alpha$, the scalar field's frequency $\omega$ and the rotational number $m$. The parity is imposed via the appropriate boundary conditions while the number of individual constituents of the chain is fixed by a suitable initial guess of the $\phi$-function. For the metric functions, the initial guess is chosen to be Minkowski. We start with a small gravitational coupling $\alpha$ together with an input value for $\omega$ close to $\omega_\text{max}$. Once a solution is obtained, $\alpha$ can be increased by small steps and we construct sequences of BSs by varying $\omega$ also by small steps.

\section{The solutions}

\label{solutions}

\begin{figure*}
    \centering
    \includegraphics[scale=1.0]{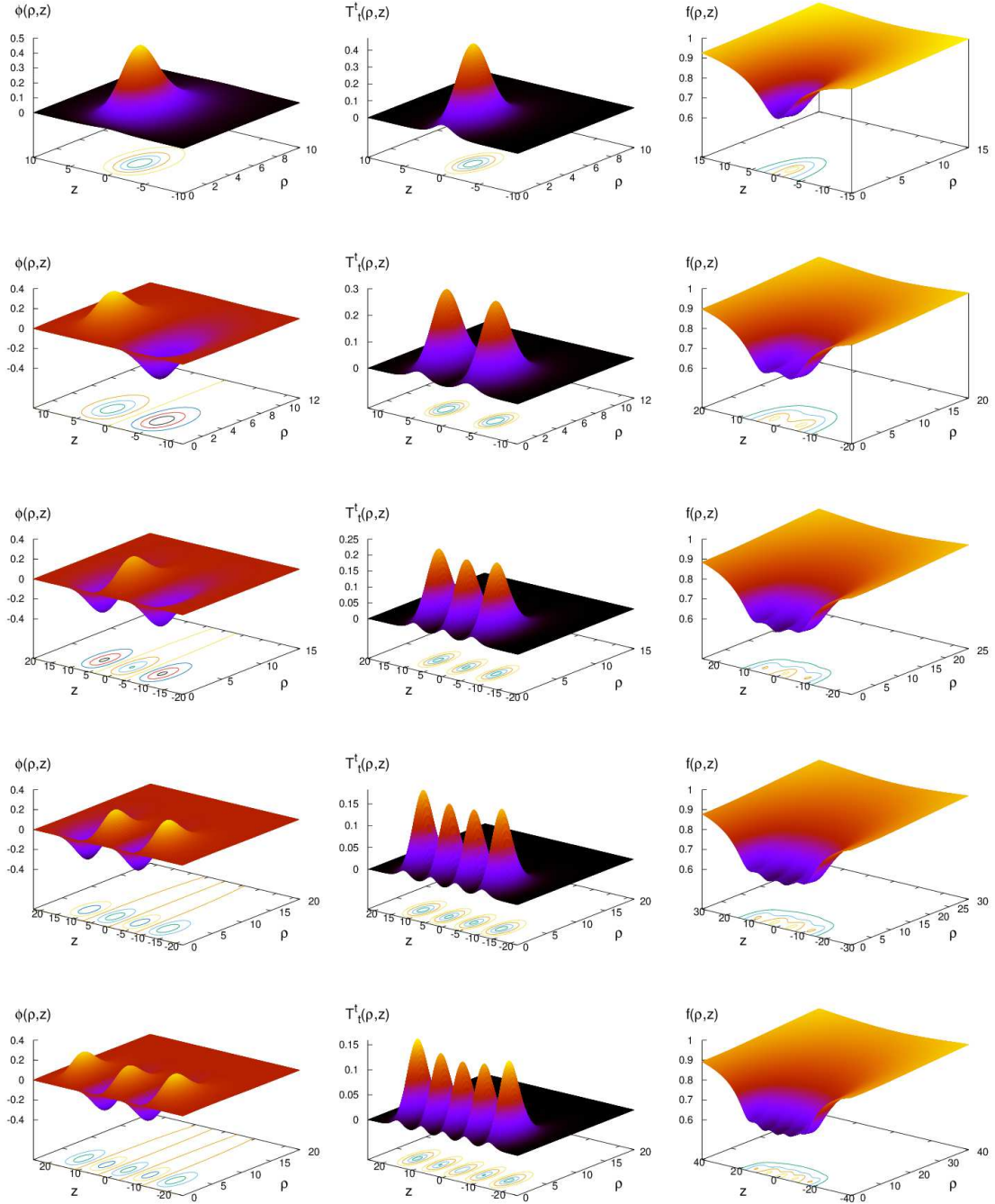}
    \caption{Chains of BSs with one to five constituents (from top to bottom) on the fundamental branch for $\alpha=0.25$, $\omega=0.9$ and $m=1$. The plots represent the scalar field amplitude $\phi$ (left), the energy density $\tensor{T}{^t_t}$ (middle) and the metric function $f$ (right) against the cylindrical coordinates $(\rho,z)$.}
    \label{main_chains}
\end{figure*}

We have constructed numerical solutions corresponding to chains of rotating BSs. Some of them had already been considered in the literature, for example, even (resp. odd) parity configurations of Ref.~\cite{Kleihaus2008} correspond to single BSs (resp. BS pairs); but we have also constructed solutions which had never been studied before.

What we call the number of constituents in the chain corresponds to the number of extrema of the scalar field amplitude $\phi$ or, equivalently, the number of maxima of the energy density. Chains of BSs are classified according to the parity of the scalar field function: chains with an even (resp. odd) number of constituents have an odd (resp. even) $\phi$-profile.

\begin{figure*}
    \centering
    \includegraphics[scale=0.65]{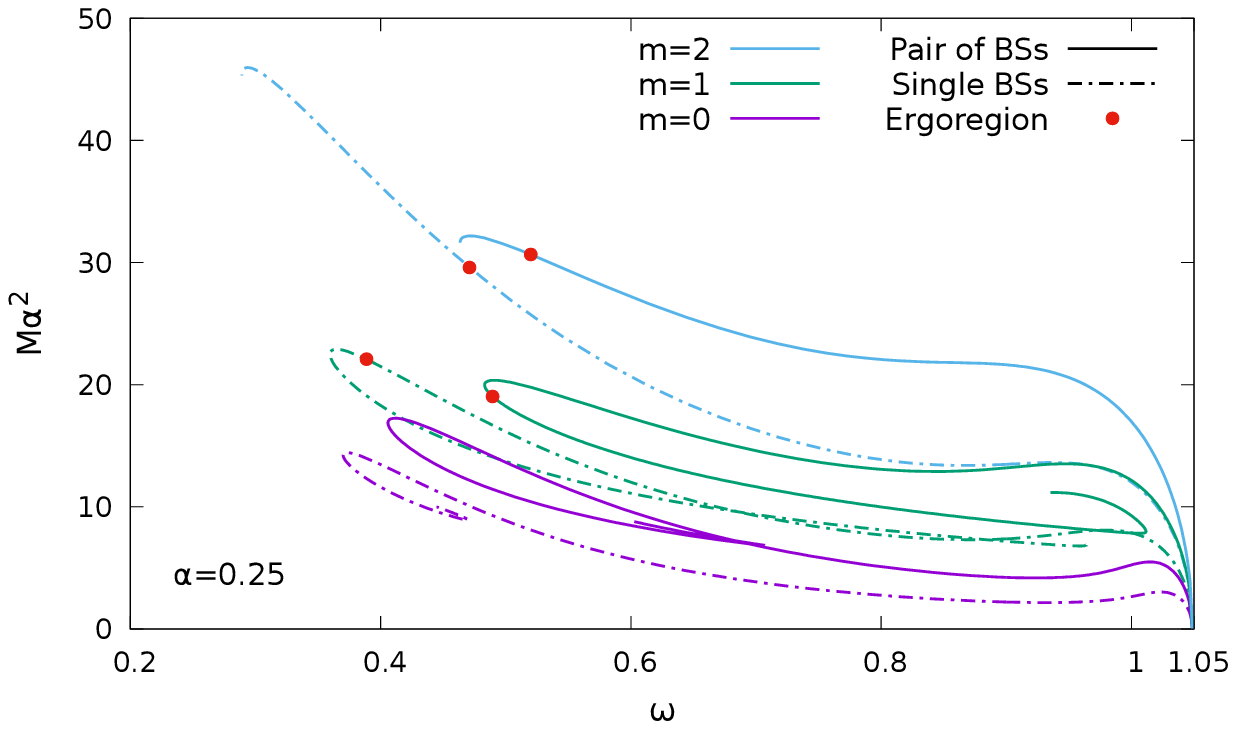}
    \includegraphics[scale=0.65]{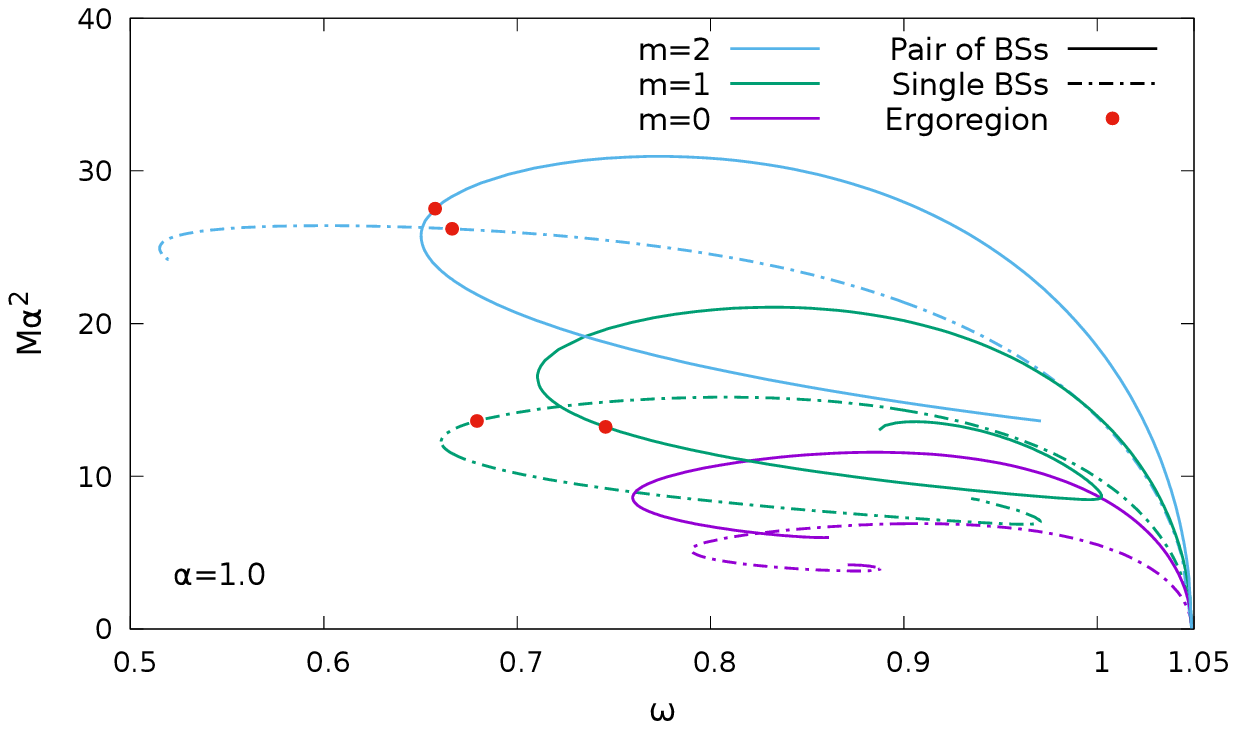}
    \includegraphics[scale=0.65]{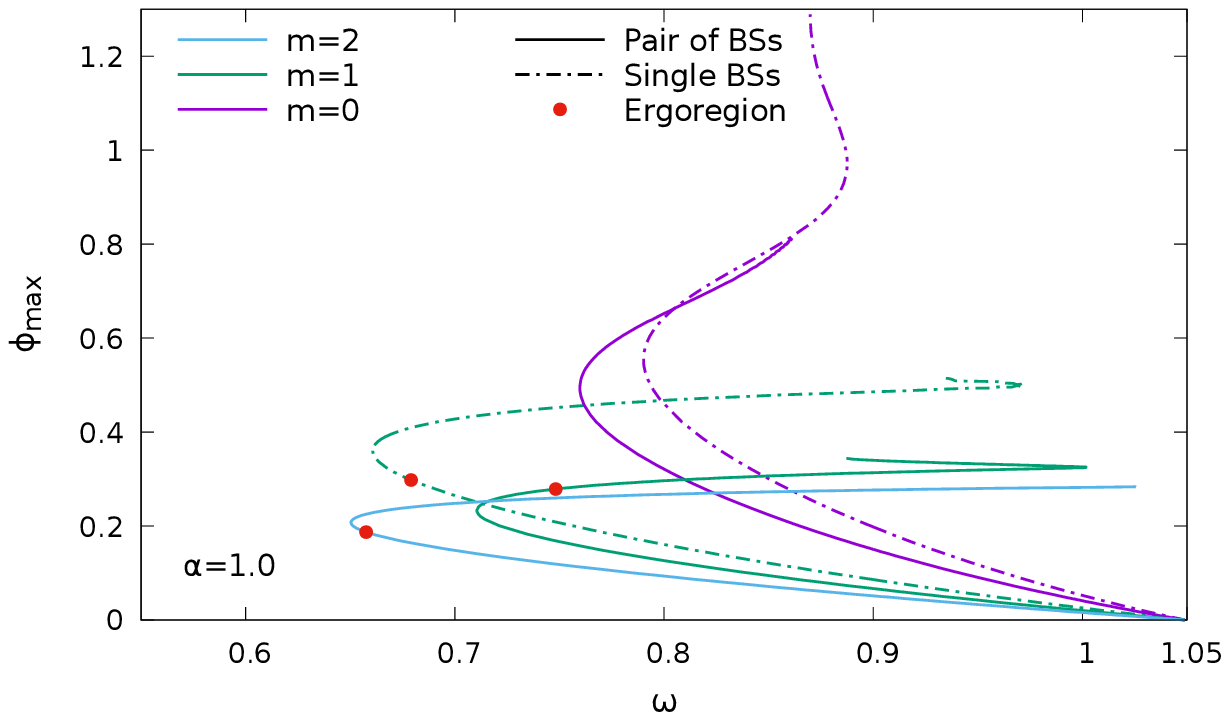}
    \includegraphics[scale=0.65]{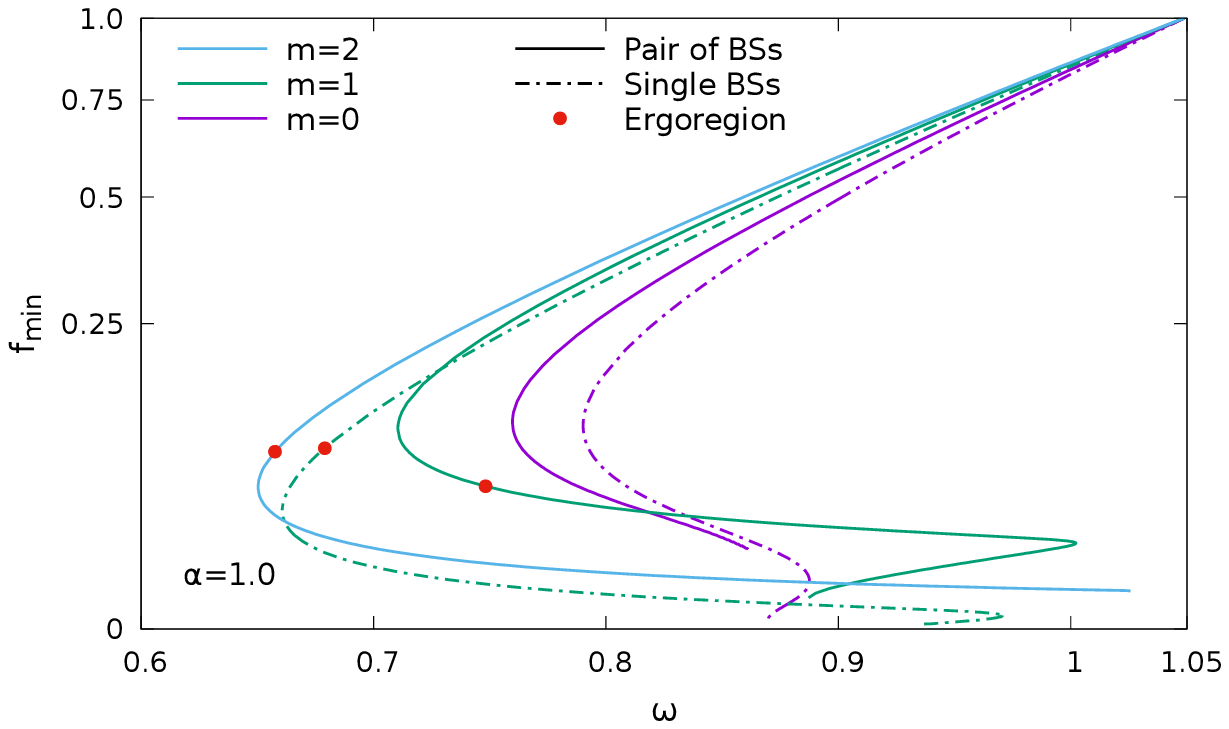}
    \caption{Scaled mass $M$ (upper panels), maximal value of the scalar field amplitude $\phi_\text{max}$ (lower left panel), minimal value of the metric function $f_\text{min}$ (lower right panel) against the frequency $\omega$ for single BSs (dash-dotted curves) and pairs (solid curves). Different values of the rotational number $m$ and gravitational coupling $\alpha$ are presented. The onset of ergoregions is indicated by red dots. The scale for $f_\text{min}$ is quadratic.}
    \label{M_om_sing_doub}
\end{figure*}

To illustrate this, we present in Fig.~\ref{main_chains} typical examples of rotating solutions where the different functions of interest, $\phi$, $\tensor{T}{^t_t}$, $f$, are plotted against the quasi-isotropic cylindrical coordinates $\rho=r\sin\theta$ and $z=r\cos\theta$. Single BSs are shown on the first row, the $\phi$-function (left column) shows a single peak -- just as the energy density $\tensor{T}{^t_t}$ does (middle column) -- and has a parity $\mathcal{P}=1$. Pairs of BSs are shown on the second row, the scalar field amplitude is now anti-symmetric, it has a parity $\mathcal{P}=-1$ with one peak and one trough. Therefore the number of extrema is two while the energy density possesses two symmetric peaks, \textit{i.e.} two maxima. We also show the lapse squared function $f$ (right column), its profile exhibits as many minima as the number of constituents in the chain and they are located where the maxima of the energy density are. For chains with a larger number of constituents, the profiles present very similar features, the $\phi$-amplitude shows alternating peaks and troughs which are related to symmetric peaks for the energy density and symmetric troughs for the lapse. Considering a surface of constant energy density gives the spatial structure of the solutions. Single BSs have a typical torus shape just like rotating $Q$-balls (see \textit{e.g.} Ref.~\cite{Radu2008}), pairs have a double tori shape, triplets correspond to triple tori and so on. All these solutions are rotating generalizations of the static chains presented in Ref.~\cite{Herdeiro2021}. 

\subsection{Single Boson Stars and pairs}

We will now recall the main results on single and pairs of BSs \cite{Kleihaus2005,Kleihaus2008}. The solutions emerge from the vacuum at the maximal value of the scalar field frequency $\omega_\text{max}=u_0$. Contrary to their flat space counterparts (rotating $Q$-balls) the mass $M$ and charge $Q$ of BSs do not diverge in this upper limit but vanish. Decreasing $\omega$ spans the first or fundamental branch of solutions; it ends at a frequency $\omega_\text{min}$ whose value depends on the gravitational coupling $\alpha$ and the rotational number $m$. Considering the upper panels of Fig.~\ref{M_om_sing_doub}, one sees that the mass $M$ remains finite when the minimal frequency is reached but the curves show their first backbending. Then the sequences of solutions can be extended by moving towards larger frequencies. Second branches of solutions start after the first backbendings, third branches after the second backbendings, \textit{etc}.

\begin{figure*}
    \centering
    \includegraphics[scale=1.0]{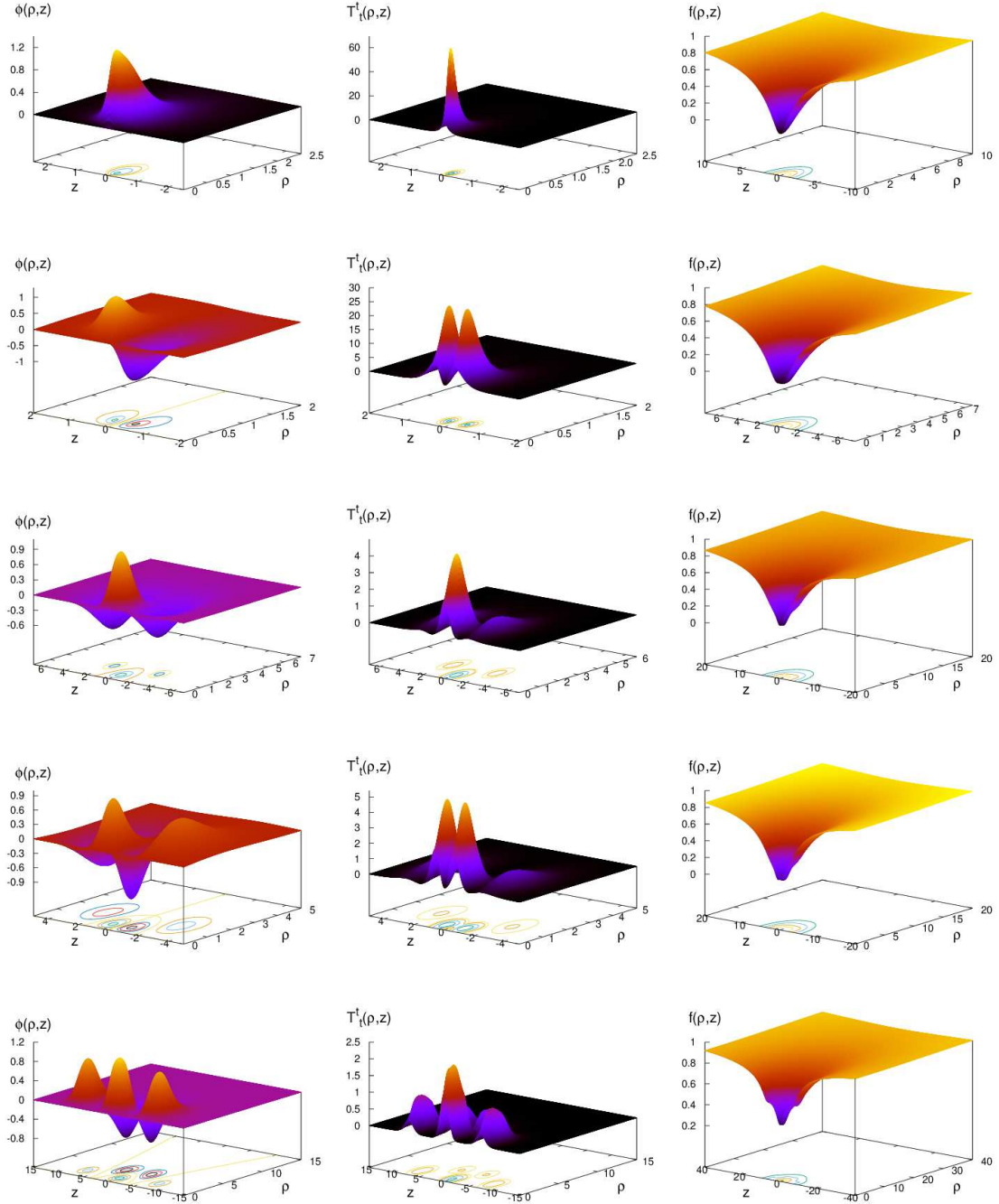}
    \caption{Chains of BSs with one to five constituents on the second branch with $m=1$, $\alpha=\{0.25,0.25,0.25,0.25,0.15\}$ and $\omega=\{0.5,0.7,0.7,0.7,0.7\}$ (frop top to bottom). The plots represent the scalar field amplitude $\phi$ (left), the energy density $\tensor{T}{^t_t}$ (middle) and the metric function $f$ (right) against the cylindrical coordinates $(\rho,z)$.}
    \label{second_chains}
\end{figure*}

The curves for single and pairs of BSs exhibit an inspiralling behaviour, approaching a limiting solution at the center of the spirals. A complete determination of spirals is however extremely difficult in general \cite{Kleihaus2005}. Therefore to avoid very time-consuming numerical computations, we only present the first few branches in our figures. If we trace $(\omega,Q)$-diagrams they would show a similar inspiralling pattern as it can be seen in Refs.~\cite{Kleihaus2005,Kleihaus2008}. Examples of profiles for solutions on the second branch can be seen in Fig.~\ref{second_chains}. For the chains with one or two constituents considered in this subsection, the main features remain unchanged but the extrema of the scalar field amplitude and the energy density are sharper and closer to the $z$-axis as compared to solutions on the fundamental branch. This means that the $\phi$-function presents in this region very high second derivatives, rendering the numerical computations more challenging.

\begin{figure*}
    \centering
    \includegraphics[scale=0.65]{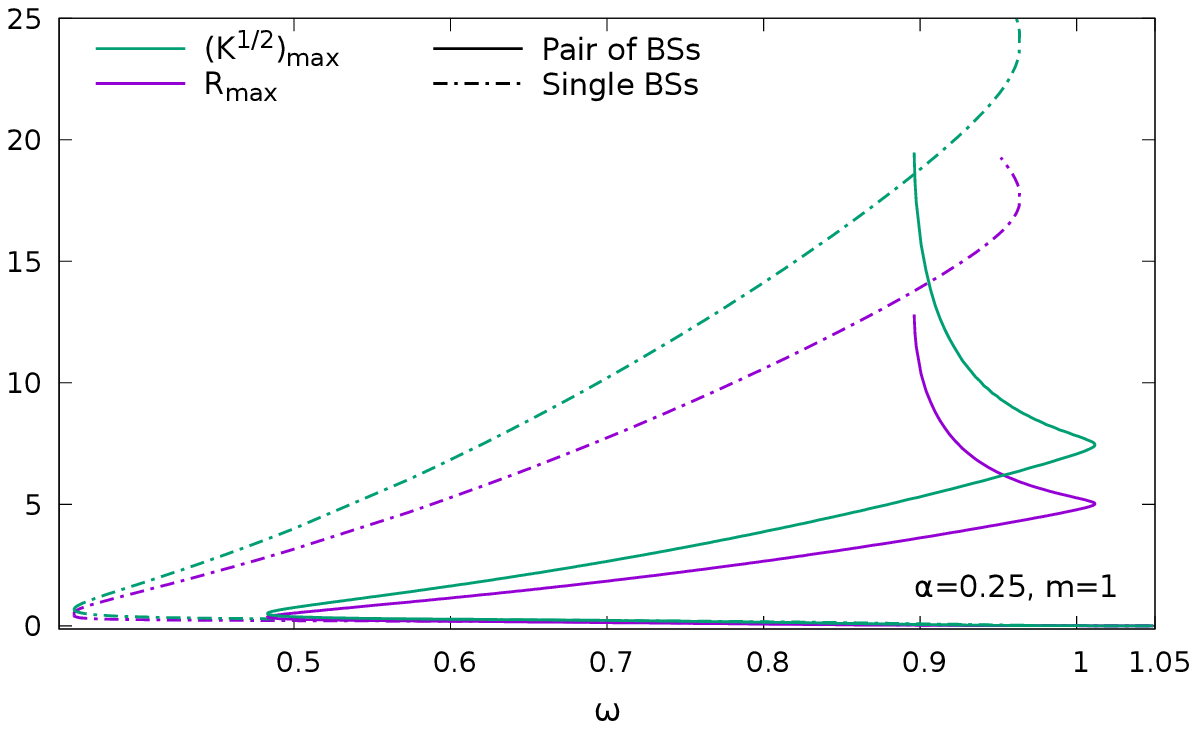}
    \includegraphics[scale=0.65]{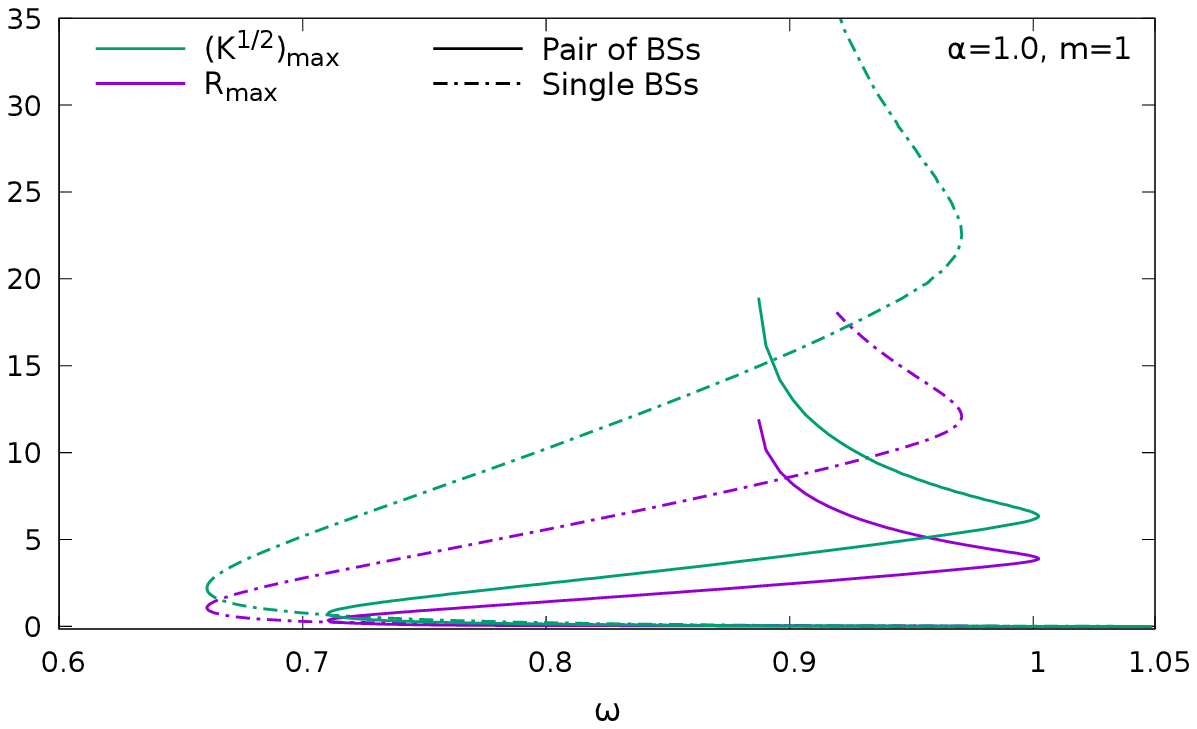}
    \caption{The maximal values of the Ricci scalar $R$ and the Kretschmann scalar $K$ against the frequency $\omega$ for single BSs (dash-dotted cuvres) and pairs (solid curves) with $m=1$ and two different values of the gravitational coupling $\alpha$.}
    \label{R_K_sing_pair}
\end{figure*}

We also present the frequency dependence of the maximal value of scalar field function $\phi_\text{max}$ (lower left panel) and the minimal value of metric function $f_\text{min}$ (lower right panel). Instead of an inspiralling behaviour, these two quantities present damped oscillations. On one hand, the maximal value of the scalar field amplitude goes from zero at $\omega_\text{max}$ when the solutions emerge and then grows as we move to the different branches. On the other hand, the minimal value of the lapse function begins from the vacuum value $\sqrt{f_\text{min}}=1$, and then approaches zero after (presumably) infinitely many oscillations.

From the comparison between the curves for $\alpha=0.25$ (upper left panel) and $\alpha=1$ (upper right panel), one can see that a larger gravitational coupling $\alpha$ increases the minimal value of the frequency $\omega_\text{min}$ and thus reduces the domain of existence of solutions. The influence of the rotational number on the value of $\omega_\text{min}$ is more complicated. Indeed, for rotating ($m\geq 1$) single BSs and pairs of BSs, increasing $m$ decreases the value of $\omega_\text{min}$; but when we pass from non-rotating ($m=0$) to rotating solutions ($m\geq1$), whether the minimal frequency increases or decreases depends on the gravitational coupling $\alpha$. For example, $\omega_{\min}$ increases for BSs pairs with $\alpha=0.25$ if we compare the $m=0$ to the $m=1$ sequence (upper left panel of Fig.~\ref{M_om_sing_doub}). In contrast, for $\alpha=1.0$ (upper right panel), $\omega_\text{min}$ decreases when $m$ goes from $0$ to $1$.

The figures also show the onset of ergoregions which occurs only when $m\geq 1$. The ergosurface is defined as the set of points such that
\begin{equation}
    g_{tt}=-f+\frac{\ell}{f}w^2\sin^2\theta=0,
\end{equation}
and the ergoregion resides inside this hypersurface. Thus in the static case, one has $w=0$ so that $g_{tt}=-f$ is always non-zero. The onset of ergoregions is indicated by red dots on the curves. Generically, an ergoregion emerges on the first or second branch and then the solutions further down the spiral all possess one. The presence of an ergoregion for regular objects like BSs is related to a superradiant instability \cite{Cardoso2008} and a light ring instability \cite{Ghosh2021}. Therefore the physical relevance of solutions which have an ergoregion is reduced.

Finally, we have also computed the Ricci scalar $R$ and the Kretschmann scalar $K=R_{\alpha\beta\mu\nu}R^{\alpha\beta\mu\nu}$ for sequences of single BSs and pairs. We present in Fig.~\ref{R_K_sing_pair} their maximal values as functions of the frequency $\omega$. This provides a new information about the limiting configuration at the center of the spirals. Indeed, the curves start from zero when the solutions emerge from the flat vacuum and then the increase of $R_\text{max}$ or $K_\text{max}$ as one moves towards the different branches becomes larger and larger. If the spirals are indeed infinite, it strongly suggests that the curvature invariants diverge and become infinite for the limiting solutions. As a result, the latter are certainly singular and thus unlikely to be numerically obtained.

\subsection{Chains with odd numbers of constituents}

\begin{figure*}
    \centering
    \includegraphics[scale=0.65]{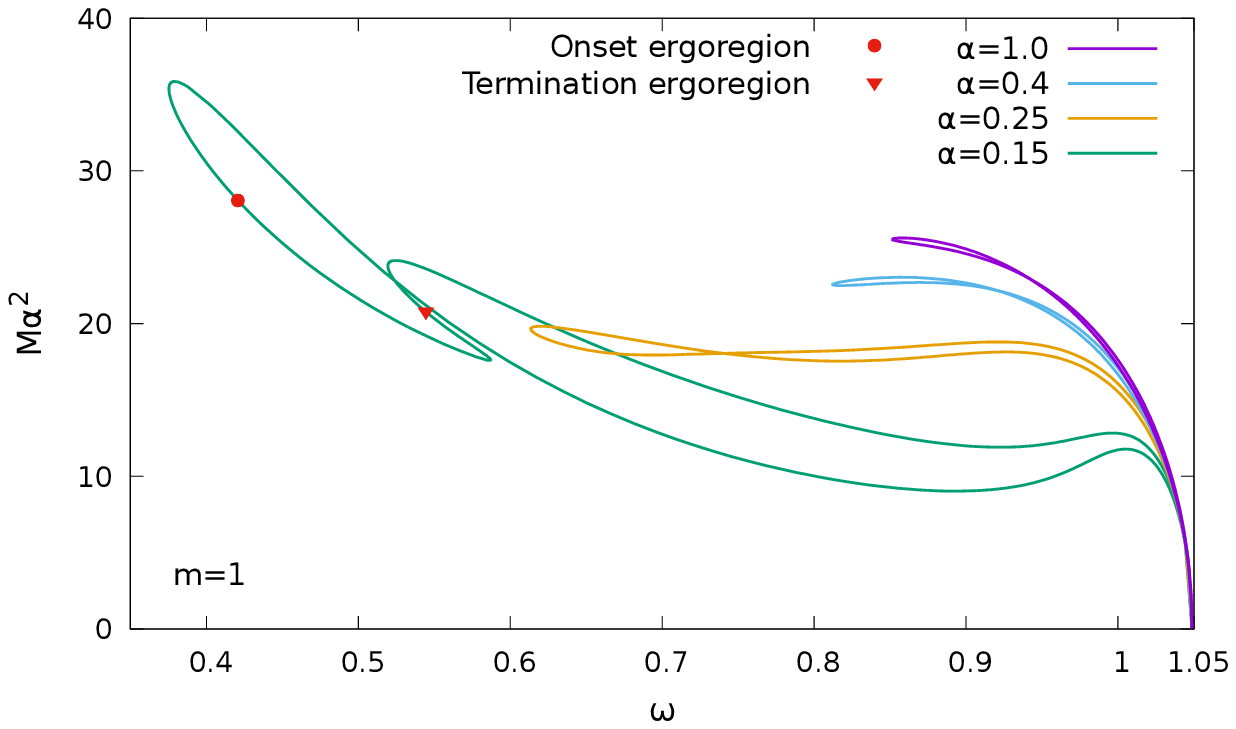}
    \includegraphics[scale=0.65]{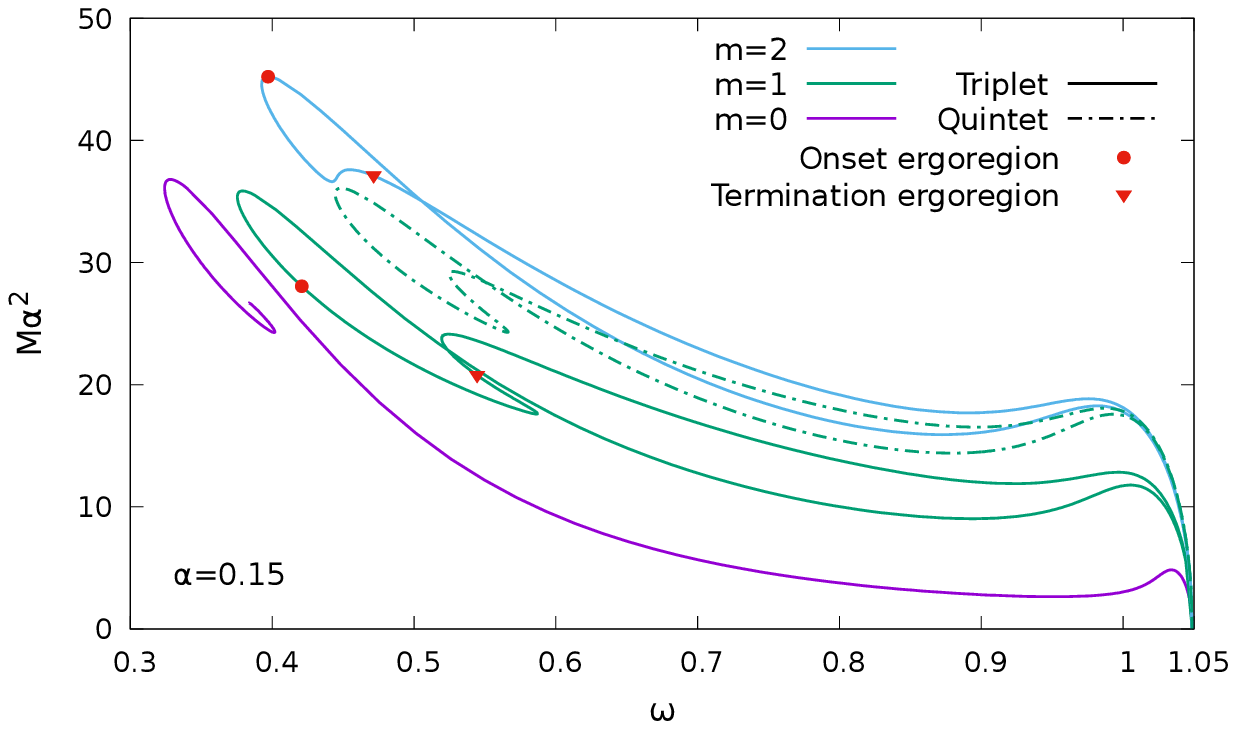}
    \includegraphics[scale=0.65]{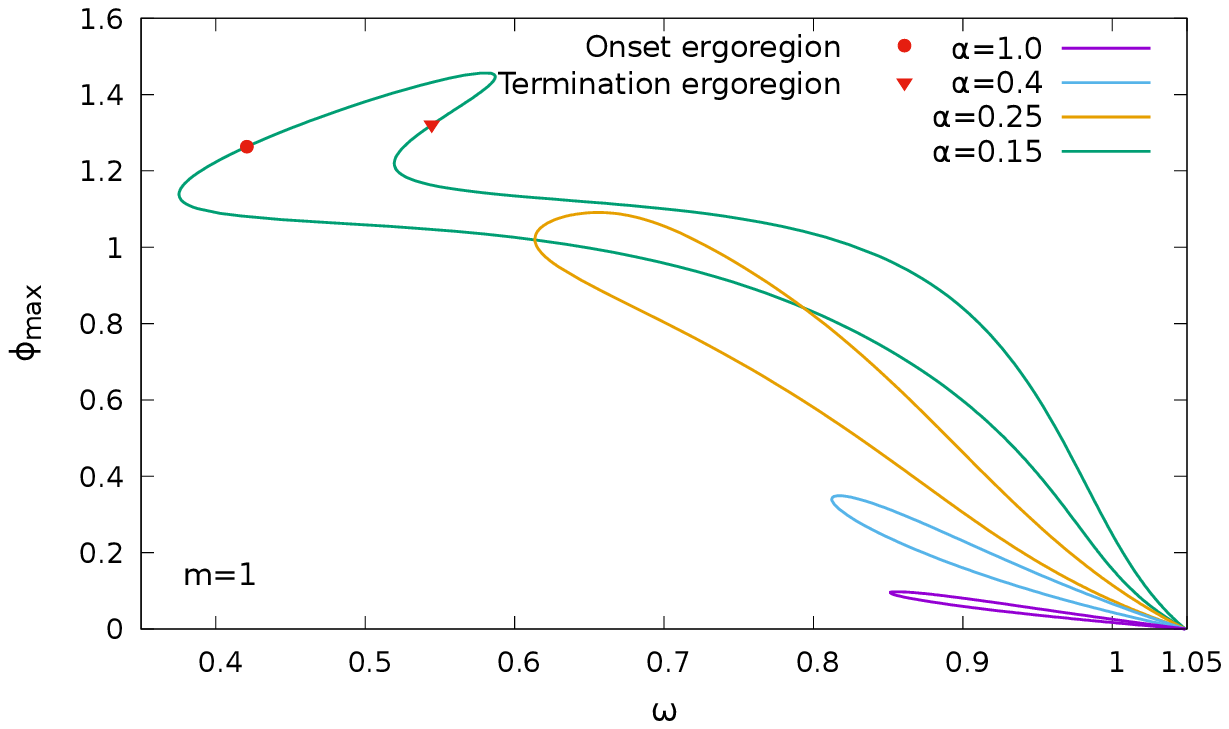}
    \includegraphics[scale=0.65]{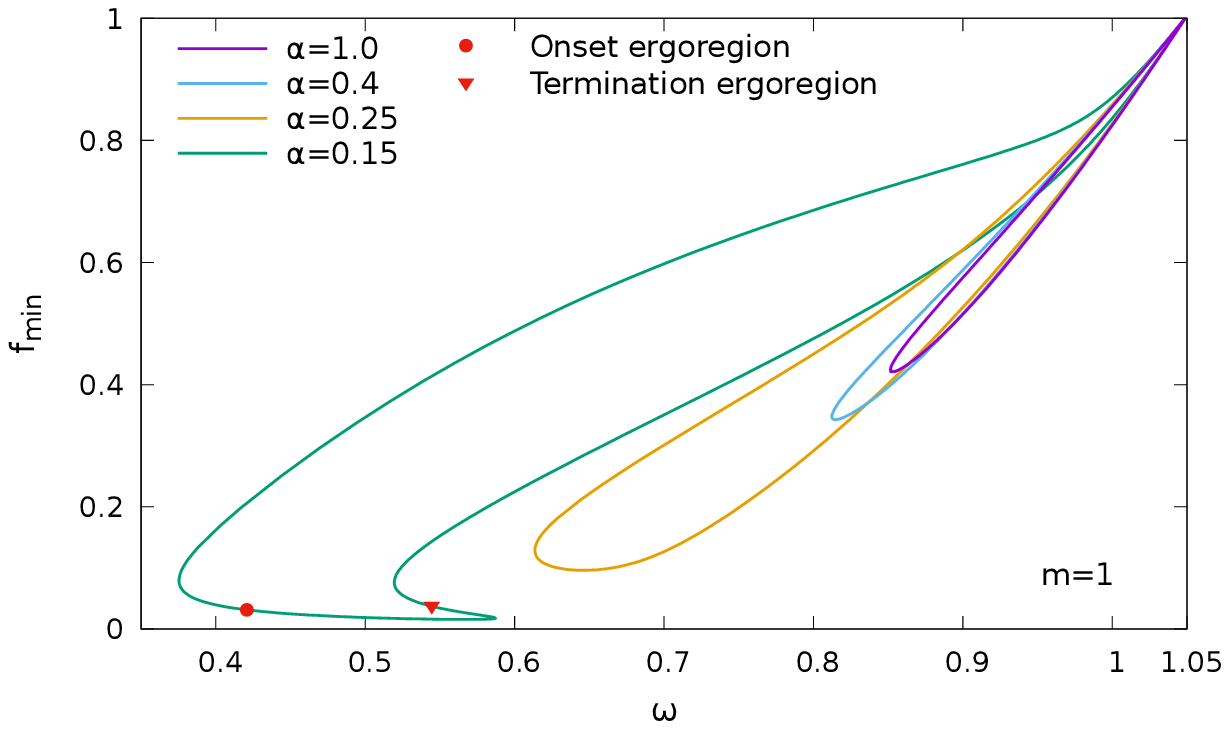}
    \includegraphics[scale=0.65]{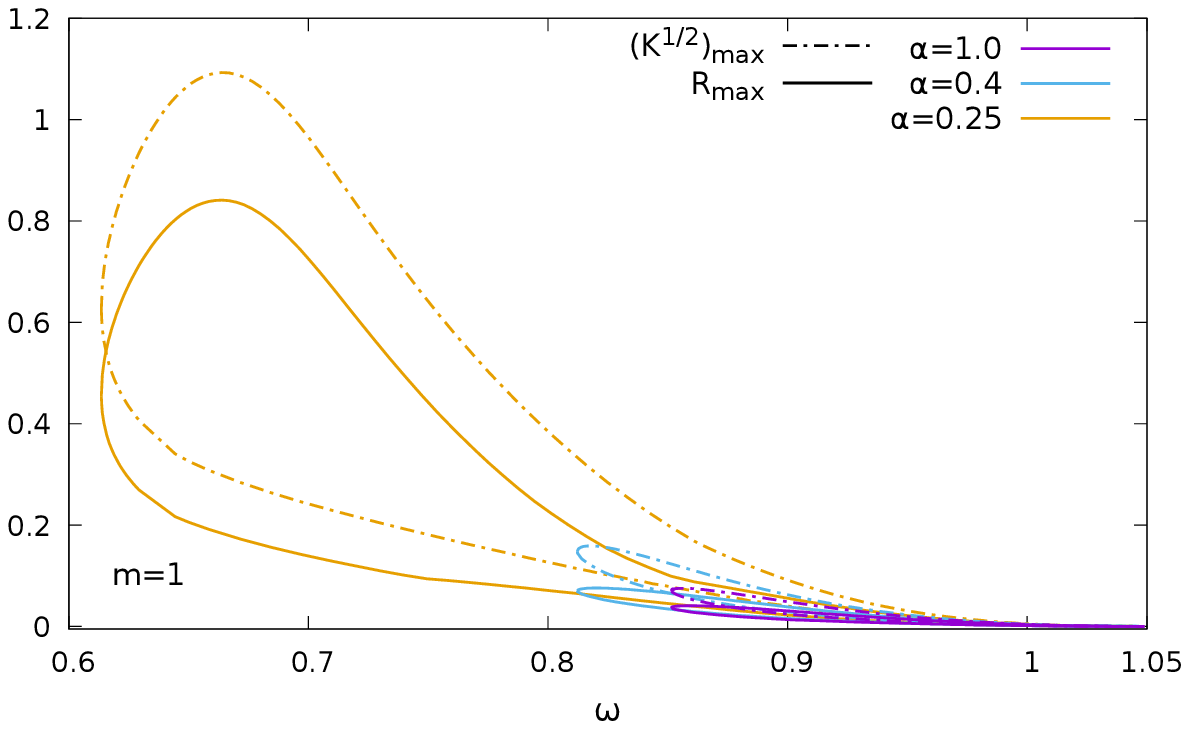}
    \caption{Triplet of BSs: the $\omega$-dependence of the scaled mass $M$ for different gravitational coupling $\alpha$ and rotational number $m$ (upper panels), maximal value of the scalar field amplitude $\phi_\text{max}$ (middle left panel), minimal value of the metric function $f_\text{min}$ (middle right panel) and maximal values of the curvature invariants $R$, $K$ (lower panel) for different values of $\alpha$ and $m=1$. The onset and termination of ergoregions are indicated respectively by red dots and triangles. In the upper right panel, the curve for BS quintets with $m=1$ is also shown for comparison.}
    \label{M_om_triplet}
\end{figure*}

Let us consider higher chains of BSs with an odd number of constituents. The scalar field amplitude of such solutions is characterized by an even parity ($\mathcal{P}=1$) with one BS being centered in the equatorial plane and the other ones located symmetrically in the upper and lower semispaces; see for example the triplet and quintet in Fig.~\ref{main_chains}. Interestingly, the properties of such chains are very different from those of single BSs. The $(\omega,M)$-diagrams for triplets are shown in Fig.~\ref{M_om_triplet}. For all rotating solutions ($m>0$) the curves no longer present the inspiralling behaviour but instead form non-trivial loops. There are only two branches of solutions: a first one starting at the maximal value of the scalar field frequency $\omega_\text{max}$, extending until the backbending is reached, and a second branch leading all the way back to the vacuum configuration. 

The denominations \textit{first} and \textit{second} branches can thus seem ambiguous in this case but one can distinguish solutions belonging to one or the other. Indeed, on the fundamental branch, when $\omega$ is close to its maximal value (high-frequency regime), the peaks and troughs of the $\phi$-profile (left column of Fig.~\ref{main_chains}) are similar in shape and located along a line parallel to the $z$-axis whereas on the second branch, the central BS dominates in amplitude and the different constituents no longer form a line. For example in the third row of Fig.~\ref{second_chains}, the two satellites are located at a larger $\rho$-coordinate than the central BS. In addition, on the second branch, the central trough of the $f$-profile overlaps with that of the satellites (right column of Fig.~\ref{second_chains}). With regards to $(\omega,M)$-diagrams, for a given value of the frequency $\omega$ close to $\omega_\text{max}$, the fundamental branch is always the one with a lower mass. This suggests that the fundamental branch in the high-frequency regime is more stable than the second branch. However for lower values of $\omega$, this mass hierarchy is inverted. 

%\caption{{\color{blue}Triplet of BSs: the $\omega$-dependence of the scaled mass $M$ for different gravitational coupling $\alpha$ and rotational number $m$ (first and second panels), maximal value of the scalar field amplitude $\phi_\text{max}$ (third panel) and minimal value of the metric function $f_\text{min}$ (fourth panel) for different values of $\alpha$ and $m=1$. The onset and termination of ergoregions are indicated respectively by red dots and triangles. In the second panel, a sequence of BSs quintet is also shown for comparison.}}

As seen in the upper left panel of Fig.~\ref{M_om_triplet}, the curves for different values of the gravitational coupling $\alpha$ present different features. Although the two branches always intersect at some value of the frequency, the curves in the low-frequency regime can present more complicated patterns when $\alpha$ is small. For example the curves for $\alpha\gtrsim 0.25$ are very similar in shape but for $\alpha=0.15$, the second branch exhibits two successive backbendings before returning to the vacuum configuration. One can also see that ergoregions do not necessarily occur: for BS triplets with rotational number $m=1$, they only appear below a critical value $\alpha_\text{cr}\lesssim 0.25$. Because of the loop structure, if there is an onset of ergoregion (red dots), a termination (red triangles) necessarily occurs: the second branch terminates at the flat vacuum configuration which obviously does not have ergoregions. Thus the solutions in the high-frequency regime should not have one neither. Finally, as for single BSs, increasing the value of $\alpha$ reduces the domain of existence of solutions. 

We have also investigated the influence of the rotational number on odd chains, see the upper right panel of Fig.~\ref{M_om_triplet} where sequences of solutions have been constructed with a fixed value of $\alpha$ and $m=0,1,2$. A larger rotational number seems to increase the minimal value of the scalar field frequency and thus reduces the domain of existence of the solutions. We also note that rotating odd chains have a different branch structure than the static ones. Indeed, for $m=0$, $\alpha=0.15$, the non-rotating triplets exhibit the same inspiralling pattern as single and pairs of BSs. In fact, as noted in Ref.~\cite{Herdeiro2021}, static triplets can have a loop structure for large gravitational coupling whereas rotating triplets seem to never show the spiral pattern, even when $\alpha$ is small. More importantly, the authors of Ref.~\cite{Herdeiro2021} had shown that when the non-rotating odd chains present the loop structure (large $\alpha$), the second branch overlaps with the fundamental branch of a spherically symmetric radially excited single BSs sequence.

\begin{figure}
    \centering
    \includegraphics[scale=0.65]{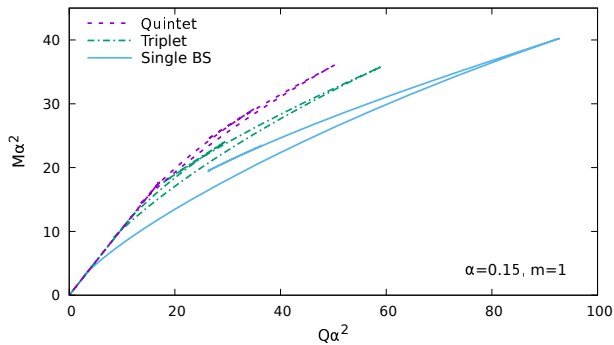}
    \caption{The scaled mass $M$ against the scaled charge $Q$ for chains of BSs with odd numbers of constituents with $m=1$ and $\alpha=0.15$.} 
    \label{M_Q_odd_chains}
\end{figure}

We expected a similar phenomenon for rotating odd chains because they present a loop structure. However it turns out that our solutions coincide exactly with excited rotating single BSs previously constructed in Ref.~\cite{Collodel2017}. On one hand, we have found no other solutions different from the radially excited BSs of Ref.\cite{Collodel2017}. On the other hand, the chain structure of the solutions profiles (see Fig.~\ref{main_chains}) is in total agreement with their non-rotating counterparts, the static chains of Ref.~\cite{Herdeiro2021}. We conclude that rotating chains with an odd number of constituents exist, and correspond to radial excitations of single BSs. To support this result, we present in Fig.~\ref{M_Q_odd_chains} the mass of single BSs, triplets and quintets versus their charge $Q$. One can see that for a given charge, the less energetic solution always lies on the sequence of single BSs. Even in the region where the different curves overlap, the single BSs sequence is still below the two others. Therefore if triplets and higher odd chains are unstable, they could possibly radiate their energy keeping their charge fixed, and decay into a single BS. To rigorously confirm such a scenario, fully time dependent simulations would be required, which is out of scope of the present paper.

For the sake of completeness, we also show in the middle and lower panels of Fig.~\ref{M_om_triplet} the maximal value of the scalar field amplitude $\phi_\text{max}$, the minimal value of the metric function $f_\text{min}$ and the maximal values of the curvature invariants $R_\text{max}$, $K_\text{max}$ against the frequency $\omega$ for BS triplets. The curves also form loops but contrary to the $(\omega,M)$-diagrams they do not intersect. It is worth noting that whereas single BSs and pairs with fixed rotational number $m$ could be uniquely parametrized by $\phi_\text{max}$, $f_\text{min}$, $R_\text{max}$ or $K_\text{max}$, this is not possible for higher odd chains because of the loop structure. This peculiarity was also noted in Ref.~\cite{Collodel2017} but with another parameter.

We have checked that this loop behaviour occurs for the BS quintets (see the upper right panel of Fig.~\ref{M_om_triplet}) and expect this scenario to represent a generic pattern for all rotating chains with a higher odd number of constituents.

\subsection{Chains with even numbers of constituents}

\begin{figure*}
    \centering
    \includegraphics[scale=0.65]{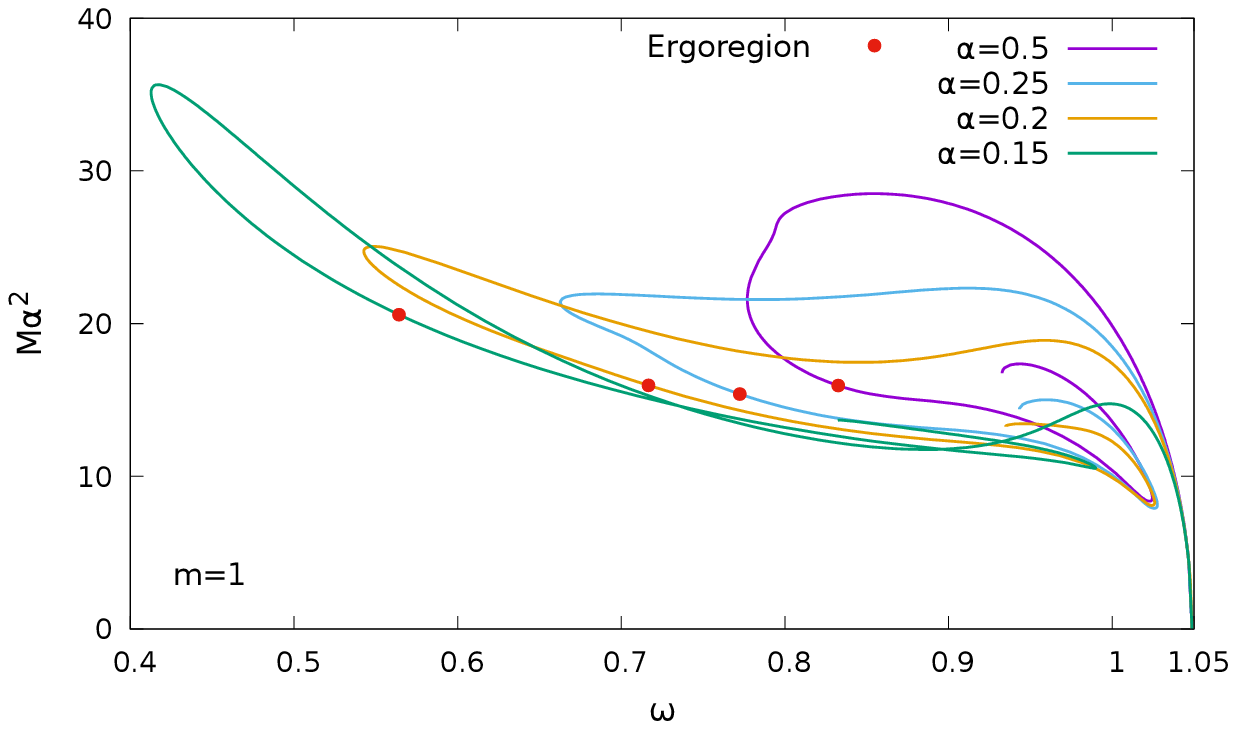}
    \includegraphics[scale=0.65]{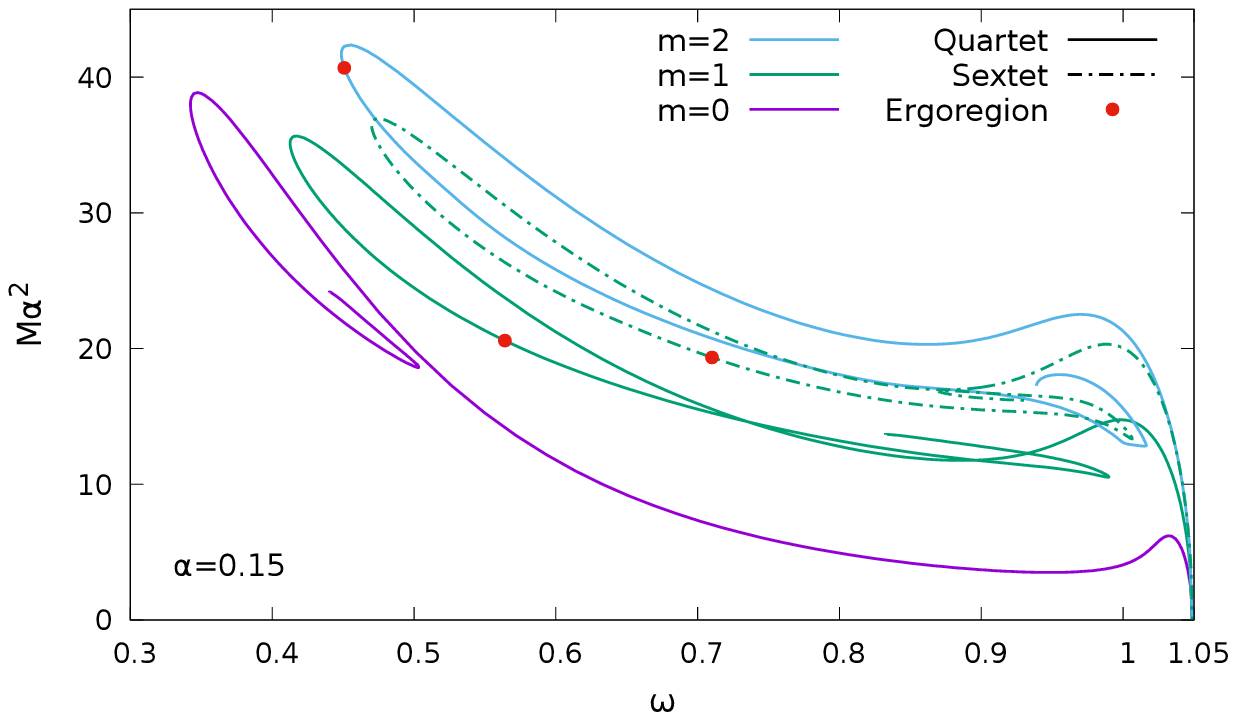}
    \includegraphics[scale=0.65]{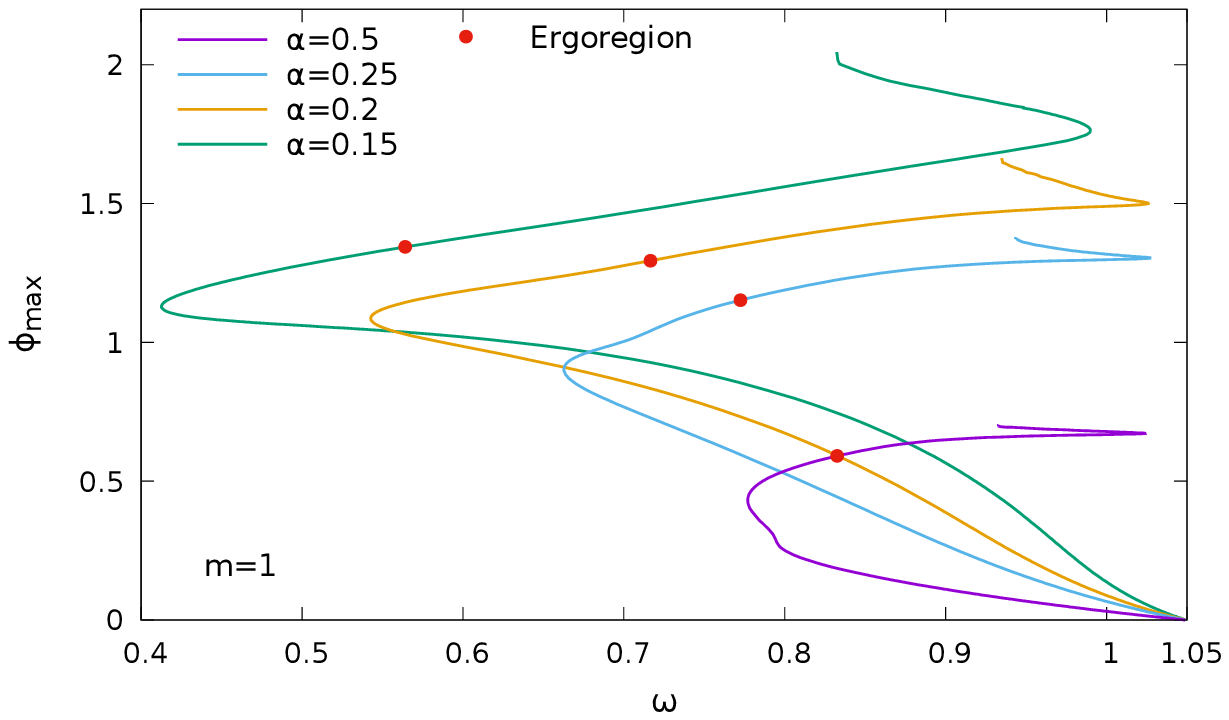}
    \includegraphics[scale=0.65]{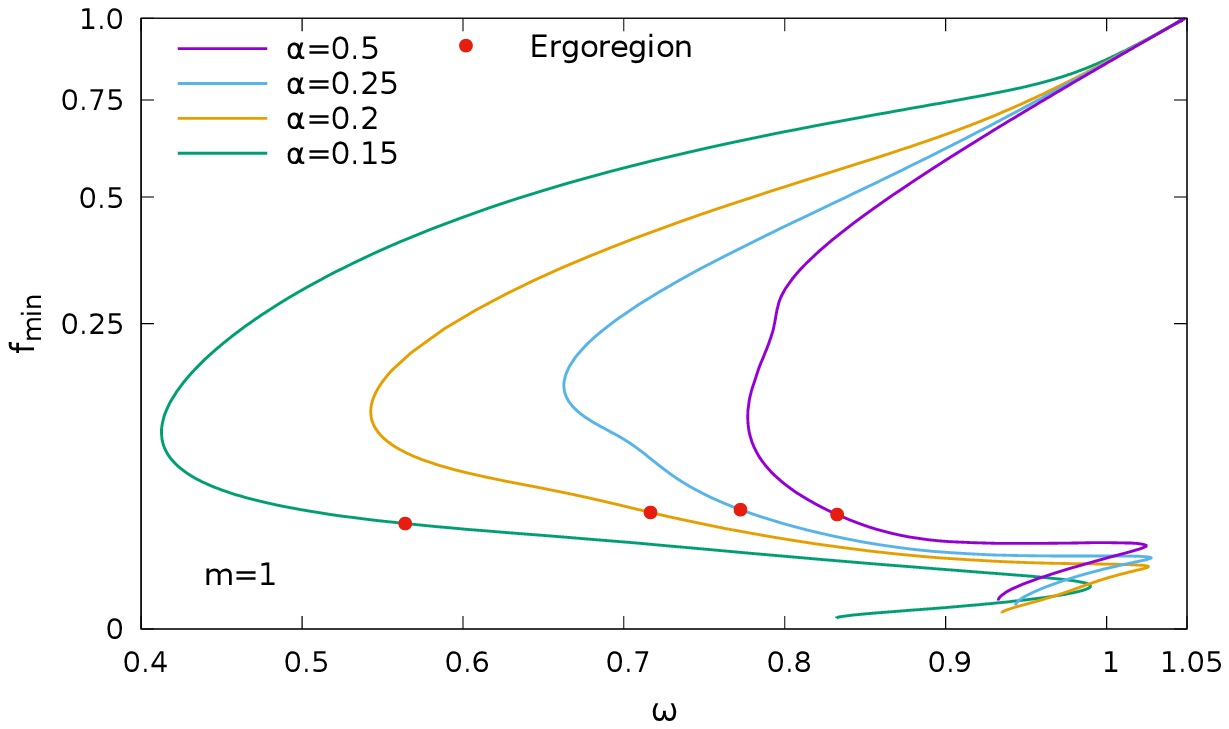}
    \includegraphics[scale=0.65]{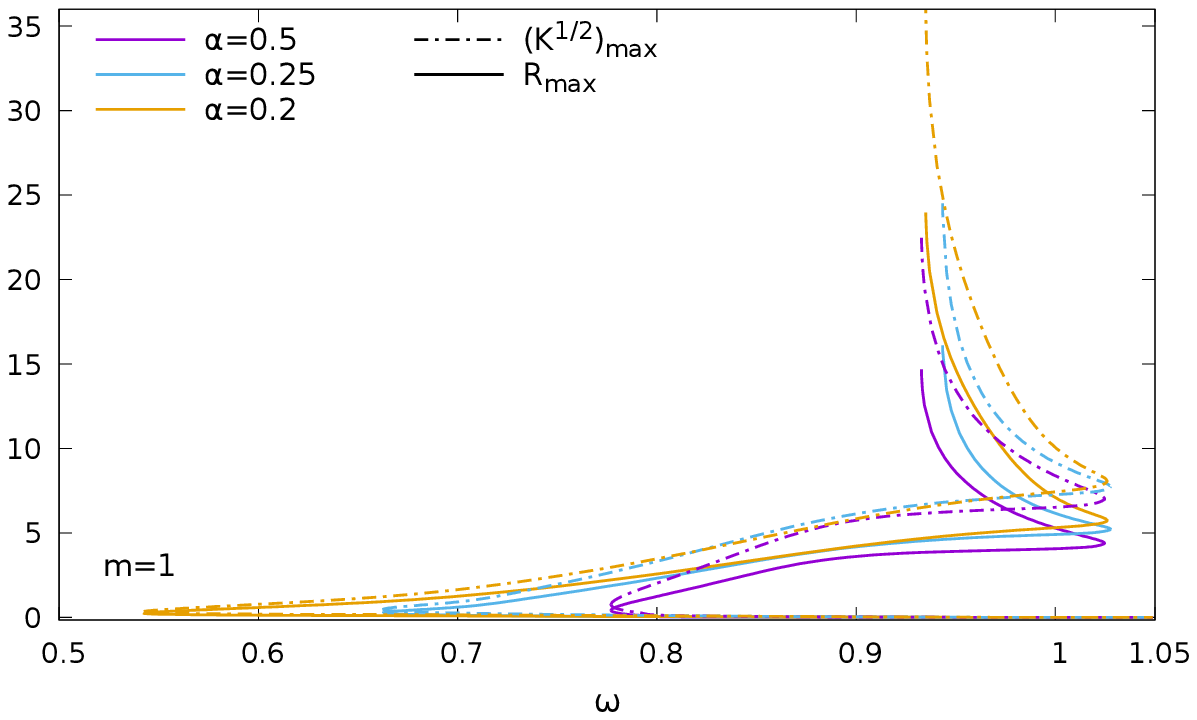}
    \caption{Quartet of BSs: the $\omega$-dependence of the scaled mass $M$ for different gravitational coupling $\alpha$ and rotational number $m$ (upper panels), maximal value of the scalar field amplitude $\phi_\text{max}$ (middle left panel), minimal value of the metric function $f_\text{min}$ (middle right panel) and maximal values of the curvature invariants $R$, $K$ (lower panel) for different values of $\alpha$ and $m=1$. The onset of ergoregions is indicated by red dots. Note the quadratic scale for $f_\text{min}$. In the upper right panel, the curve for BS sextets with $m=1$ is also shown for comparison.}
    \label{M_om_quartet}
\end{figure*}

We now turn to the rotating chains with a higher even number of constituents. Such solutions are characterized by the odd parity $\mathcal{P}=-1$ of their $\phi$-amplitude just as BS pairs. Although they are very natural generalizations of the previously obtained configurations \cite{Kleihaus2008,Herdeiro2021}, it is the first time that such solutions are explicitly constructed. 
Whereas the triplets and higher odd chains have a different branch structure than single BSs, the $(\omega,M)$-diagrams for even chains are all similar to those for BS pairs. Indeed, one can see in the upper panels of Fig.~\ref{M_om_quartet} the inspiralling behaviour with (presumably) infinitely many branches. In the upper left panel, the rotational number is fixed ($m=1$) and we consider different values for the gravitational coupling. Again, larger values for $\alpha$ yields a smaller domain of existence of the solutions. Like rotating pairs, the sequences present an ergoregion onset (red dots) and all solutions located further down the spiral possess one.

At the same time, the higher even chains of BSs share also some properties with the odd chains. For example, in the upper right of Fig.~\ref{M_om_quartet} we show sequences of quartets with $\alpha=0.15$, $m=0,1,2$ and one can see that solutions with a larger rotational number have a smaller domain of existence just as the triplets of Fig.~\ref{M_om_triplet}. Another similarity with the odd chains is the profile of the solutions as we move on the second branch. Configurations of the fundamental branch begin with all constituents aligned and parallel to the $z$-axis (see fourth row of Fig.~\ref{main_chains}) whereas on the second branch, the central BS pair dominates in amplitude and the satellites are located at a larger $\rho$-coordinate than the two central constituents (see fourth row of Fig.~\ref{second_chains}). Further in the sequence, the situation between odd and even configurations becomes different. For odd chains, the amplitude of extrema globally decreases as the solutions dissolve to the trivial vacuum, see for example the middle left panel of Fig.~\ref{M_om_triplet} where the maximal value $\phi_\text{max}$ of the scalar field amplitude for triplets is shown. In contrast for even chains, only the amplitude of the outer satellites decreases while the central pair continues to grow in amplitude as we move to the different higher branches, see the middle left panel of Fig.~\ref{M_om_quartet} where $\phi_\text{max}$ \textit{vs.} $\omega$ is shown for quartets. 
%This inverted effect of the rotational number for higher chains as compared to single or pairs of BSs was never pointed out in the literature.

Unfortunately, we cannot conclude on the finiteness of the maximal value of the scalar field amplitude $\phi_\text{max}$ at the center of spirals since the central pair of the $\phi$-profile becomes extremely sharp, rendering the numerical computation very challenging. Regarding the minimal value of the metric function $f_\text{min}$, one can see on the middle right panel of Fig.~\ref{M_om_quartet} that it approaches zero as we move towards more involved spirals.

Even if it is unclear weather the scalar field function remains finite or not as approaching the center of spirals, the limiting solutions are certainly singular. This can be inferred from the lower panel of Fig.~\ref{M_om_quartet} where the maximal values of curvature invariants $R$ and $K$ are presented as functions of the frequency $\omega$. These values increase as one moves towards the different branches, and the increase itself becomes larger and larger, suggesting that the curvature invariants diverge for the limiting configurations.

\begin{figure}
    \centering
    \includegraphics[scale=0.65]{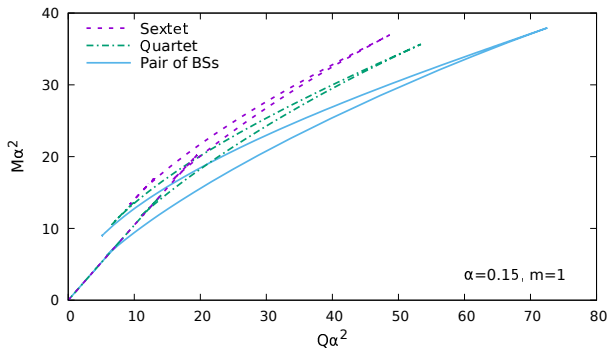}
    \caption{The scaled mass $M$ against the scaled charge $Q$ for chains of BSs with even numbers of constituents with $m=1$ and $\alpha=0.15$.}
    \label{M_Q_even_chains}
\end{figure}

We expect the spirals to occur for all even chains, although so far we have checked it only for quartets and sextets.

Finally we conjecture that the chains with a higher even number of constituents also coincide with excitations of BS pairs just as the chains with a higher odd number of constituents are radial excitations of single BSs. In support of this conjecture, we show in Fig.~\ref{M_Q_even_chains} the mass of pairs of BSs, quartets and sextets as a function of the charge $Q$. For all solutions with an even number of constituents greater than or equal to four (quartets and sextets on the figure) there is a pair configuration with the same charge and lower mass. As for the odd chains of Fig.~\ref{M_Q_odd_chains}, in the region where the curves overlap, the sequence of BS pairs is still below the two others. Therefore, chains with a higher even number of constituents can decay into a pair of BSs keeping their charge fixed. As a result, the higher even chains are probably unstable and can consistently be seen as excitations of BS pairs.

\subsection{Flat space limit}

Let us finally consider the flat space limit of the rotating chains of BSs as the gravitational coupling $\alpha$ approaches zero. Constructing full sequences of rotating solutions with small values of $\alpha$ turns out to be numerically very challenging \cite{Kleihaus2005,Kleihaus2008}: only a part of the fundamental branch can be easily constructed. We have thus fixed the value of the boson frequency $\omega$ by choosing a solution on the fundamental branch, then varied only $\alpha$. We expect our results to apply for every BS chains belonging to the fundamental branches.

\begin{figure}
    \centering
    \includegraphics[scale=0.65]{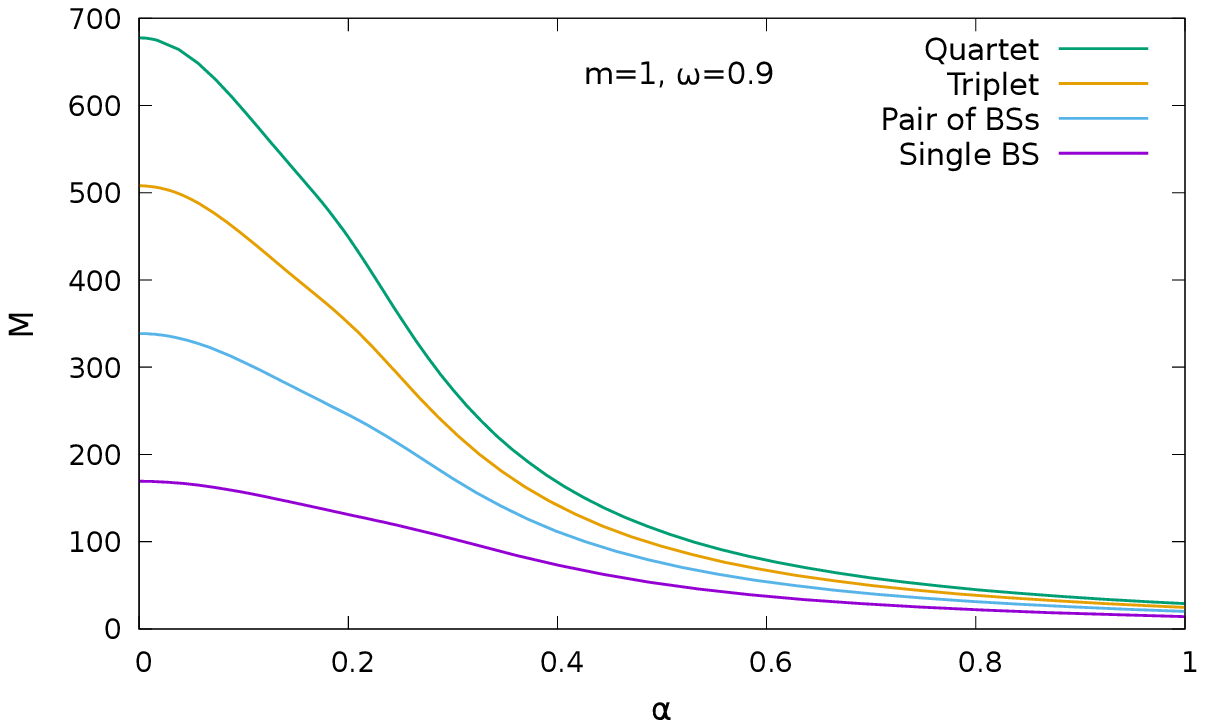}
    \includegraphics[scale=0.65]{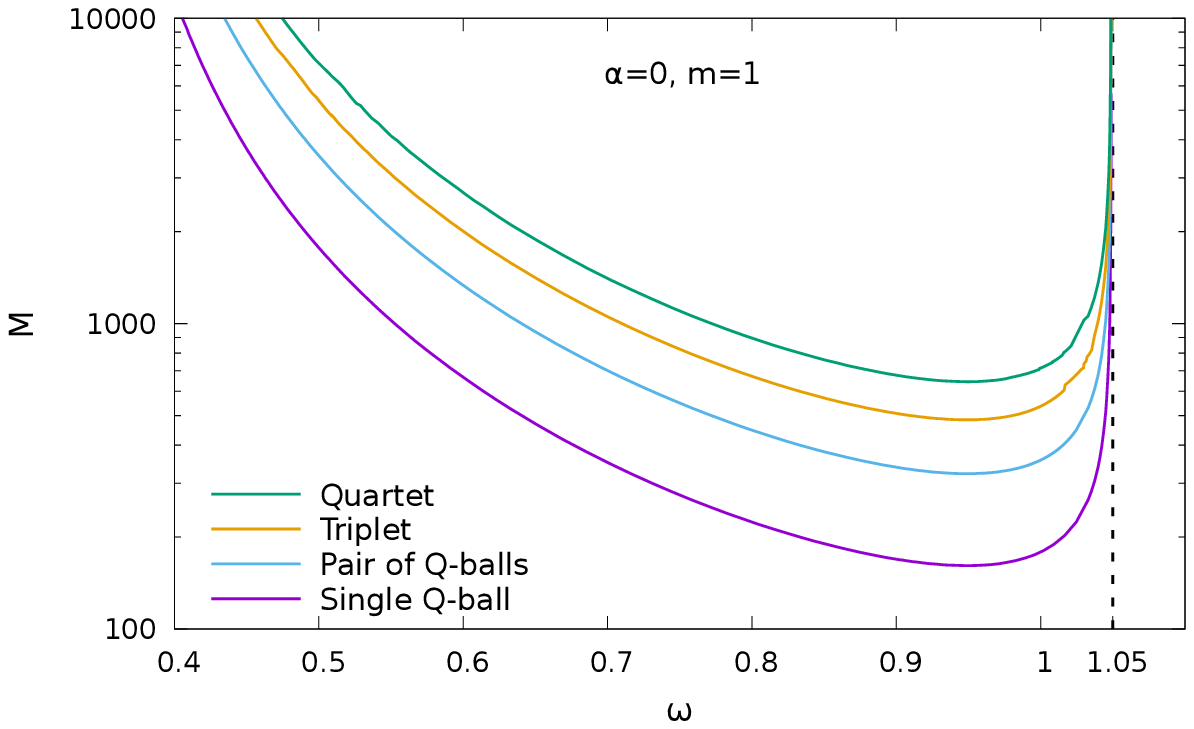}
    \caption{Upper panel: The mass $M$ of BS chains with one to four constituents on the fundamental branch against the gravitational coupling $\alpha$ for $\omega=0.9$ and $m=1$. Lower panel: The mass $M$ against the frequency $\omega$ for the chains of $Q$-balls with one to four constituents, $m=1$.}
    \label{M_al}
\end{figure}

In the upper panel of Fig.~\ref{M_al}, we present the mass $M$ of the chains with one to four constituents as a function of the gravitational coupling $\alpha$. It appears that limiting solutions exist when $\alpha\rightarrow 0$. For single BSs and pairs, these solutions are the rotating $Q$-balls with even and odd parity obtained in Refs.~\cite{Volkov2002a,Kleihaus2005,Kleihaus2008}. More interestingly, the triplets and quartets also have $Q$-balls counterpart in the absence of gravity: the limiting solutions thus correspond to chains of $Q$-balls (also referred to as $Q$-chains). To our knowledge, such configurations have never been reported in the literature although chains of non-rotating $Q$-balls minimally coupled to a U(1) gauge field have been constructed recently \cite{Loiko2021}. 

\begin{figure}
    \centering
    \includegraphics[scale=0.6]{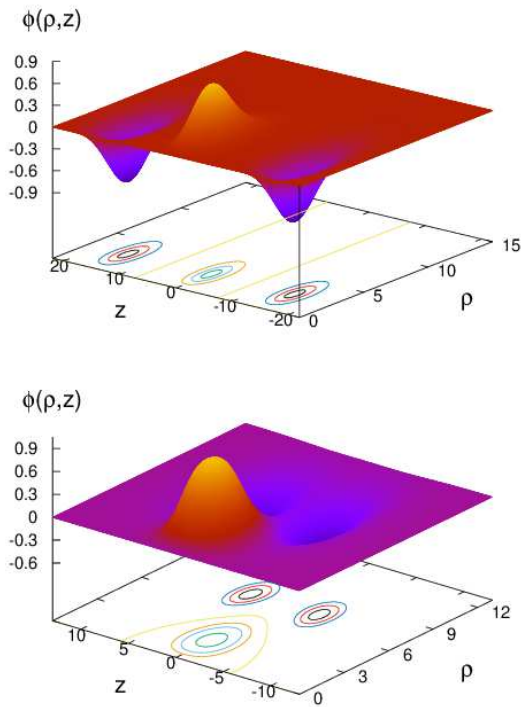}
    \caption{$Q$-balls triplets of the first family (upper panel) and second family (lower panel) with $\omega=0.9$ and $m=1$. The scalar field function $\phi$ is plotted against the cylindrical coordinates $(\rho,z)$.}
    \label{Q_balls}
\end{figure}

The profile of the scalar field function $\phi$ for the $Q$-chains is very similar to those of BSs, see the upper panel of Fig.~\ref{Q_balls}. By increasing the resolution of the mesh used for the discretization, we have checked that the chains of $Q$-balls are numerically stable: increasing the number of grid points for a given solution does not change the $\phi$-profile and in particular, the distance between neighboring constituents remains unchanged.

We construct the sequences of $Q$-chains by varying the frequency $\omega$ and present in the lower panel of Fig.~\ref{M_al} their mass $M$ as a function of $\omega$. The frequency dependence of the charge $Q$ is qualitatively similar. The main features already known for single $Q$-balls or pairs seem to hold also for chains with higher numbers of constituents. The mass and the charge diverge in the upper limit of the frequency domain at $\omega=\omega_\text{max}$ where $\omega_\text{max}$ is still given by Eq.~\eqref{ommax}. The domain is also bounded from below at a finite value of the frequency determined uniquely in terms of the potential parameters \cite{Volkov2002a,Kleihaus2005,Kleihaus2008}
\begin{equation}
    \omega_\text{min}^\text{$Q$-ball}=\sqrt{u_0^2-\frac{\lambda^2}{4}}.
\end{equation}
Mass and charge diverge as well in this limit. It is worth noting that contrary to the BSs case, the different constituents of these $Q$-chain remain aligned throughout the sequence.

Up to the numerical accuracy, the mass of a $Q$-chain with $n$ constituents coincides with $n$ times the mass of a single $Q$-ball. This is not very surprising since in the absence of gravity, the different constituents interact together only via the scalar interaction which is short-ranged. Furthermore, the distance between neighboring $Q$-balls in a chain is sufficiently large to ensure that they almost do not interact and as a consequence, the energy of the whole $Q$-chain is very close to the the sum of the energies of the different constituents taken individually.

It remains to clarify how the higher branches of BS chains evolve in the flat space limit. For the single BSs, this analysis has already been carried out by \citet{Kleihaus2005}. They found that the spiral get shifted to the low frequencies and that the minimal frequency for the BSs becomes smaller than the one for $Q$-balls as $\alpha$ approaches zero. Therefore for very small values of $\alpha$, the BSs belonging to higher branches in the spiral have their frequency in the range 
\begin{equation}
    \omega\in\left[\omega_\text{min}(\alpha\rightarrow 0),\omega_\text{min}^\text{$Q$-ball}\right],
\end{equation}
and they do not have flat space limit with finite charge or mass.

For the rotating chains of BSs with even numbers of constituents, the $(\omega,M)$-diagrams also exhibit the spiraling pattern. Although we have not been able to construct full sequences for small values of the gravitational coupling, we can nevertheless conjecture that the higher branches of BS chains with even numbers of constituents do not have a regular flat space limit just like the single BSs.

\begin{figure}[b!]
    \centering
    \includegraphics[scale=0.65]{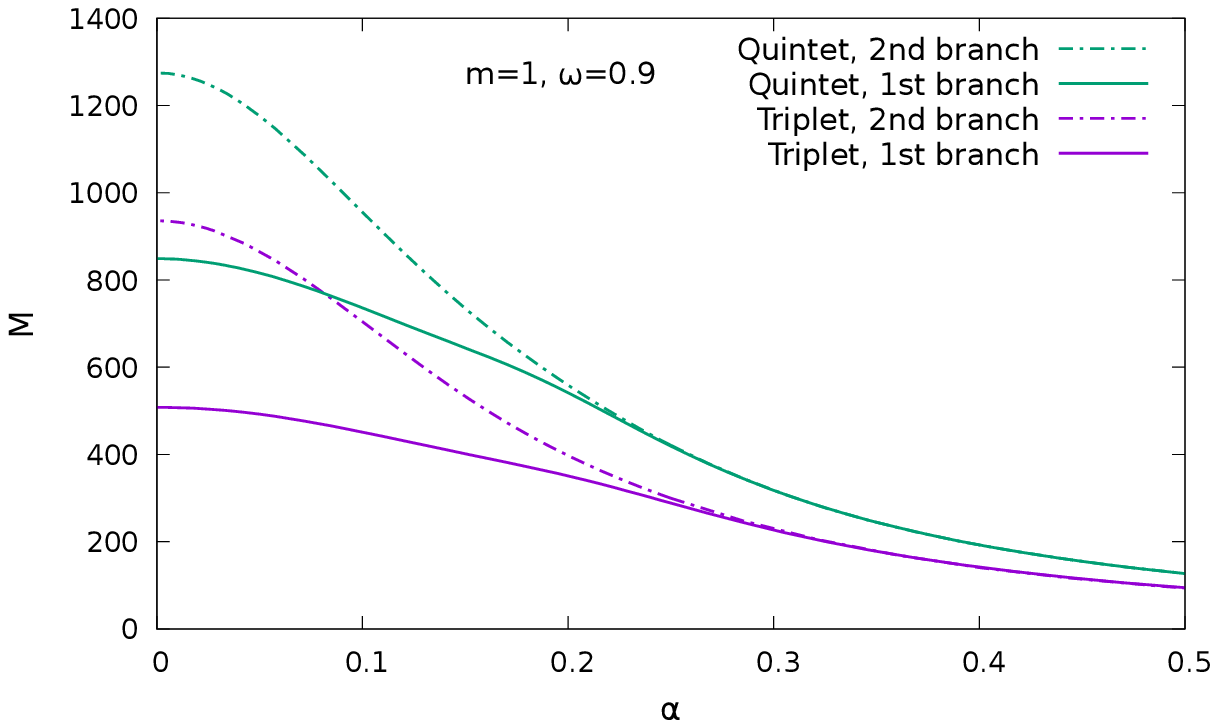}
    \includegraphics[scale=0.65]{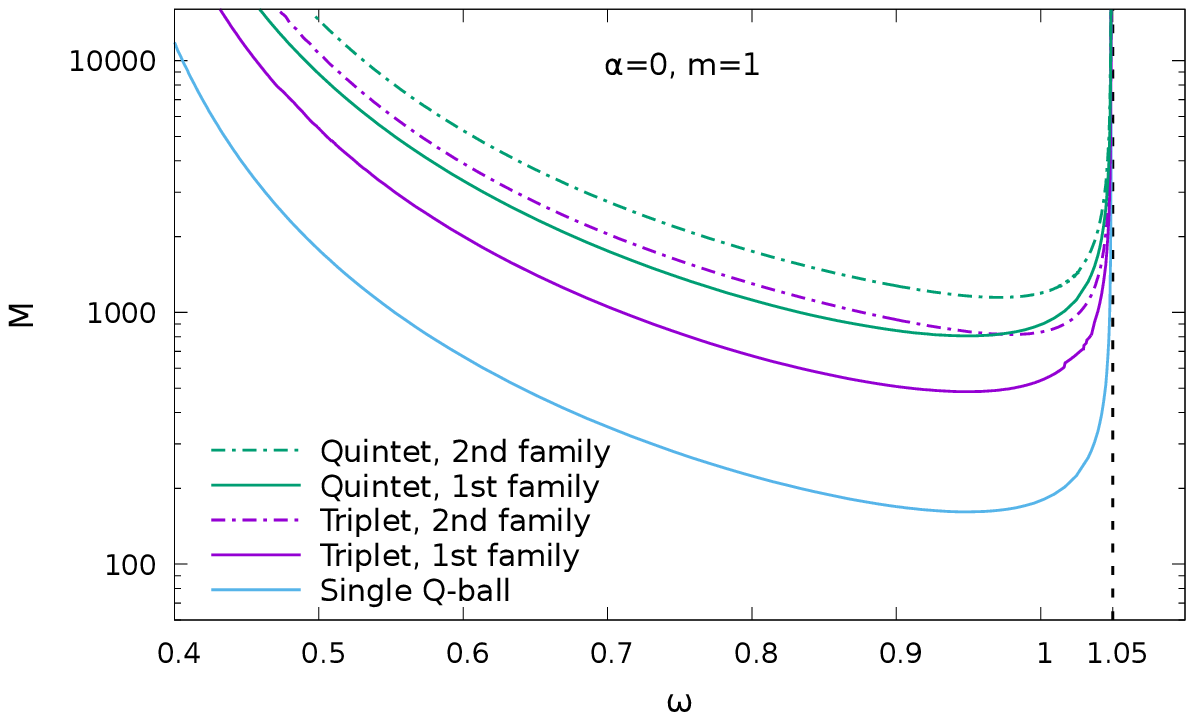}
    \caption{Upper panel: The mass $M$ of BS triplets and quintets on the first and second branches against the gravitational coupling $\alpha$ for $\omega=0.9$ and $m=1$. Lower panel: The mass $M$ against the frequency $\omega$ for the two families of $Q$-balls triplets and quintets with $m=1$. For comparison, the curve for single $Q$-balls is also shown.}
    \label{M_al_ex}
\end{figure}

We now turn to the chains of BSs with an odd number of constituents. Such configurations have their second branch starting at $\omega_\text{min}$ and ending at $\omega_\text{max}$ where the solutions converge to the flat vacuum. Choosing a solution on the second branch and fixing the value of $\omega$, we are able to find the limiting solution for vanishing gravitational coupling. We present the $\alpha$ dependence of the mass for BS triplets and quintets on the second branch and compare to the curves for the fundamental branch in the upper panel of Fig.~\ref{M_al_ex}. Interestingly, the two flat space solutions are different, indicating the existence of new families of $Q$-balls. 

We construct the sequences of these new families of solutions for triplets and quintets and show them in the lower panel of Fig.~\ref{M_al_ex}. For every frequency the second families are more energetic than the first ones. A typical example of $\phi$-profile for a triplet is shown in the lower panel of Fig.~\ref{Q_balls}: a central constituent is centered in the equatorial plane and is surrounded by the other companions but the different constituents are not aligned. We can see that the two satellites are close to the central $Q$-ball and thus the interaction between them is no longer negligible, leading to a high energy of the global configuration as compared to the first family.

As a result, the set of rotating solutions in the $Q$-ball theory is much richer than expected. There is still an open problem to check whether the $Q$-chains are stable or not. If the chains of BSs are unstable as we conjecture, it seems very unlikely that their flat space counterpart would be stable in the absence of an attractive interaction provided by gravity.

We expect our results to generalize for higher numbers of constituents and for all rotational number $m\geq 1$. However the flat space limit of the non-rotating ($m=0$) chains of BSs found by \citet{Herdeiro2021} remains an open issue. We have tried the same strategy as in the rotating case to find the limiting solutions for $\alpha\rightarrow 0$ but we obtain numerically unstable configurations: increasing the grid resolution leads to an increase of distances between the neighboring constituents.

\section{Conclusion and perspectives}

\label{conclusion}

We have addressed the rotating generalization of the static chains of BSs reported recently in Ref.~\cite{Herdeiro2021}. They are solitonic configurations of a self-interacting complex scalar field minimally coupled to Einstein gravity which possess a harmonic time dependence governed by the boson frequency $\omega$ and a non-zero angular momentum. Similar chains were also studied in Ref.~\cite{Herdeiro2021a} but for a scalar field without self-interactions. Since the single rotating BSs have a toroidal shape \cite{Kleihaus2005}, the chains of BSs we have considered here consist of a stack of multiple tori.

The scalar field amplitude changes sign between two neighboring constituents in a chain. This yields a repulsive scalar interaction between the neighboring BSs \cite{Battye2000,Bowcock2009}. For an equilibrium configuration to exist, the repulsion between the different constituents has to be balanced by an attractive interaction. In the present case, this ingredient for the existence of chains is provided by the gravitational attraction.

We have numerically constructed the rotating chains of BSs by using the finite element method implemented in the FreeFem++ solver. To test the robustness of our code, we have reproduced already known solutions: rotating single BSs, pairs \cite{Kleihaus2005,Kleihaus2008} and non-rotating chains \cite{Herdeiro2021}. Convergence tests have been carried out and they show a fourth order convergence in the number of points used for the discretization. To the best of our knowledge, it is the first time this finite element software is used for gravitational problems and we are confident that it can be used for more sophisticated models.

We have found rotating chains of BSs with up to six constituents but they are likely to exist for an arbitrarily large number of constituents. The solutions exist in a finite frequency range just as the single BSs or the pairs. The upper bound is completely fixed by the mass of the scalar excitations around the vacuum whereas the lower bound depends on the strength of the gravitational coupling $\alpha$ and on the rotational number $m$ present in the harmonic azimuthal dependence of the scalar field. We have constructed sequences of solutions with fixed $\alpha$, $m$ and focused our interest on the frequency dependence of parameters such as the mass, the maximal value of the scalar field function or the maximal values of curvature invariants. We have also discussed a qualitative argument regarding the stability of the chains and analyzed their flat space limit as the gravitational coupling approaches zero.

On the one hand, the sequences of chains with an \textit{odd} number of constituents do not present a spiraling pattern when their mass is considered as a function of the frequency, as single BSs do. Instead, the $(\omega,M)$-diagrams form nontrivial loops starting and ending at the flat vacuum. The consequence is that such solutions cannot be uniquely parametrized by a single parameter. It turns out that the BS triplets were previously obtained in the literature \cite{Collodel2017} and the configurations were referred to as radially excited BSs. We emphasize that the starting point of the present work was the non-rotating chains of BSs considered in Ref.~\cite{Herdeiro2021}. Only after constructing the BS triplets, we discovered that they correspond to the solutions of Ref.~\cite{Collodel2017}. As a result, all the chains with an odd number of constituents may correspond to excitations of a single BS. To illustrate this, we have shown that for a fixed Noether charge, triplets and quintets are more energetic than a single BS. We expect this mass hierarchy to hold also for all chains with a higher odd number of constituents. Because the charge is related to the number of particles in a configuration and has to be conserved, triplets, quintets and other higher odd chains are likely to decay into a single BS with the same number of particles or equivalently, with the same charge. When the gravitational coupling approaches zero, the odd chains of BSs reduce to two different families of $Q$-balls chains which have never been reported in the literature. 

On the other hand, the sequences of chains with an \textit{even} number of constituents present a spiraling frequency dependence of their mass and an oscillating pattern when the maximal value of the scalar field, the minimal value of the lapse or the maximal values of the curvature invariants are considered as functions of the frequency. Furthermore, the evolution of curvature invariants along the sequences indicates that the limiting solutions at the center of spirals are certainly singular. We have also found chains of $Q$-balls with even numbers of constituents corresponding to the flat space limit of even BS chains. Although we have checked these properties only for pairs, quartets and sextets, we expect our results to be generic for all chains with a higher even number of constituents. Similarly as for the odd chains, we have shown that for a given charge, the quartet and the sextet are more energetic than the fundamental BS pair and are thus likely to be unstable. As a result, we conjecture that all the chains with a higher even number of constituents correspond to excitations of BS pairs.

A dynamical study of the chains of BSs is still lacking, both to confirm the possible decay of the chains with a number of constituents greater than or equal to three into single BSs or pairs and to find potential scenarios for the formation of these objects. It is worth noting that the dynamical evolution of such a non-linear physical system is certainly very complex and cannot be inferred from the present work. Although some numerical time evolution and stability analysis of BSs with a self-interacting potential have already been performed previously \cite{Becerril2007,Kleihaus2012,Liebling2017,Siemonsen2021}, they only concern non-rotating or rotating \textit{single} BSs \footnote{Dynamical evolutions of \textit{mini}-BSs can be found in the literature \cite{Palenzuela2007,SanchisGual2019} but for such objects, the scalar field potential contains only a mass term.}. In any case, the onset of ergoregions in the sequences of BS chains indicates the presence of an instability for the configurations which possess one.

\section*{Acknowledgments}

The author gratefully acknowledges E. Radu and Y. Shnir for useful discussions, L. Suleiman and L. Villain for their careful proofreading and finally J. Garaud for his introduction to FreeFem++.

\appendix

\begin{widetext}
\section*{Appendix: System of Partial Differential Equations}

\label{expliciteq}

The set of coupled elliptic PDEs we obtain after injecting the ansatz \eqref{ansatzmet}, \eqref{ansatzphi} to the field equations is the following
\begin{align*}
    r^2\phi_{,rr}+\phi_{,\theta\theta}+2r\,\phi_{,r}+\text{cot}\,\theta\,\phi_{,\theta}=&-\frac{h}{f^2}\left(\ell(r\,\omega+m\,w)^2-r^2\ell\,f\,U'(\phi^2)-f^2m^2\text{csc}^2\theta\right)\phi\\
    \label{eqphi}
    &-\frac{1}{2\ell}\left(r^2\,\ell_{,r}\phi_{,r}+\ell_{,\theta}\phi_{,\theta}\right),\numberthis{}\\
    r^2 f_{,rr}+f_{,\theta\theta}+2r\,f_{,r}+\text{cot}\,\theta\,f_{,\theta}=&\frac{\ell}{f}\big(8\alpha^2 h\left(r\,\omega+m\,w\right)^2\phi^2-4r^2\alpha^2f\,h\,U(\phi^2)\\
    &+\sin^2\theta(w-r\,w_{,r})^2+\sin^2\theta\,w_{,\theta}^2\big)\\
    &+\frac{1}{f}\left(r^2 f_{,r}^2+f_{,\theta}^2\right)-\frac{1}{2\ell}\left(r^2\ell_{,r}f_{,r}+\ell_{,\theta}f_{,\theta}\right),\numberthis{}\\
    r^2 \ell_{,rr}+\ell_{,\theta\theta}+3r\,\ell_{,r}+2\,\text{cot}\,\theta\,\ell_{,\theta}=&\frac{8\alpha^2\ell^2h}{f^2}\left((r\,\omega+m\,w)^2\phi^2-r^2f\,U(\phi^2)\right)-8\alpha^2 m^2\text{csc}^2\theta\,\ell\,h\,\phi^2\\
    &+\frac{1}{2\ell}\left(r^2\ell_{,r}^2+\ell_{,\theta}^2\right),\numberthis{}\\
    r^2 w_{,rr}+w_{,\theta\theta}+2r\,w_{,r}+3\,\text{cot}\,\theta\,w_{,\theta}=&8\alpha^2\text{csc}^2\theta\,m\,h(r\,\omega+m\,w)\phi^2+2w-\frac{2}{f}\big(r\,f_{,r}(w-r\,w_{,r})\\
    &-f_{,\theta}w_{,\theta}\big)+\frac{3}{2\ell}\left(r\,\ell_{,r}(w-r\,w_{,r})-\ell_{,\theta}w_{,\theta}\right),\numberthis{}\\
    r^2 h_{,rr}+h_{,\theta\theta}+r\,h_{,r}=&\frac{4\alpha^2 h^2}{f^2}\left(r^2f\,\ell\,U(\phi^2)+\left(3m^2\text{csc}^2\theta\,f^2-\ell(r\,\omega+m\,w)^2\right)\phi^2\right)\\
    &-4\alpha^2 h\left(r^2\phi_{,r}^2+\phi_{,\theta}^2\right)+\frac{2h}{\ell}\left(r\,\ell_{,r}+\text{cot}\,\theta\,\ell_{,\theta}\right)\\
    &-\frac{h}{2f^2}\left(r^2 f_{,r}^2+f_{,\theta}^2\right)+\frac{3h\,\ell}{2f^2}\sin^2\theta\left((w-r\,w_{,r})^2+w_{,\theta}^2\right)\\
    \label{eqh}
    &+\frac{h}{2\ell^2}\left(r^2 \ell_{,r}^2+\ell_{,\theta}^2\right)+\frac{1}{h}\left(r^2 h_{,r}^2+h_{,\theta}^2\right),\numberthis{}
\end{align*}
where we have introduced the compact notation $\phi_{,\mu}\equiv\partial_\mu\phi$.

If we fix the metric to be Minkowski $f=\ell=h=1$, $w=0$ and set to zero the gravitational coupling $\alpha$, only one PDE remains, Eq.~\eqref{eqphi}, and it describes rotating $Q$-balls.
\end{widetext}

%\bibliography{biblio_phys}


\end{document}